\documentclass[12pt]{article}
\usepackage{amsmath}
\usepackage{amsfonts}
\usepackage{amssymb}
\usepackage{mathptmx}
\usepackage{slashed}
\usepackage{latexsym}
\usepackage{graphicx}
\usepackage{color}

\newcommand{\newc}{\newcommand}
\newc{\be}{\begin{equation}}
\newc{\ee}{\end{equation}}
\newc{\bea}{\begin{eqnarray}}
\newc{\eea}{\end{eqnarray}}
\newc{\ol}{\overline}
\newc{\wt}{\widetilde}
\newc{\bs}{\boldsymbol}
\newc{\m}{\mathcal}
\newc{\ra}{\rightarrow}
\newc{\lra}{\leftrightarrow}
\newc{\ba}{\begin{eqnarray}}
\newc{\ea}{\end{eqnarray}}
\newc{\pa}{\partial}
\newc{\D}{\Delta}

\newc{\nn}{\nonumber}

\def\beq{\begin{equation}}
\def\eeq{\end{equation}}
\def\bea{\begin{eqnarray}}
\def\eea{\end{eqnarray}}

\newcommand{\pr}{\sp\prime}
\newcommand{\ov}{\overline}

\begin{document}

\begin{titlepage}

\vspace*{0.7cm}

\begin{center}
{
\bf\LARGE
Towards a Realistic F-theory GUT}
\\[12mm]
James~C.~Callaghan$^{\star}$
\footnote{E-mail: \texttt{James.Callaghan@soton.ac.uk}},
Stephen~F.~King$^{\star}$
\footnote{E-mail: \texttt{king@soton.ac.uk}},
George~K.~Leontaris$^{\dagger}$
\footnote{E-mail: \texttt{leonta@uoi.gr}},
Graham~G.~Ross$^{\S}$
\footnote{E-mail: \texttt{g.ross1@physics.ox.ac.uk}},
\\[-2mm]

\end{center}
\vspace*{0.50cm}
\centerline{$^{\star}$ \it
School of Physics and Astronomy, University of Southampton,}
\centerline{\it
SO17 1BJ Southampton, United Kingdom }
\vspace*{0.2cm}
\centerline{$^{\dagger}$ \it
Physics Department, Theory Division, Ioannina University,}
\centerline{\it
GR-45110 Ioannina, Greece}
\vspace*{0.2cm}
\centerline{$^{\S}$ \it
Rudolf Peierls Centre for Theoretical Physics,}
\centerline{\it
University of Oxford, 1 Keble Road, Oxford, OX1 3NP, UK}
\vspace*{1.20cm}

\begin{abstract}
\noindent
We consider semi-local F-theory GUTs arising from
a single $E_8$ point of local enhancement, leading to simple GUT groups
based on $E_6$, $SO(10)$ and $SU(5)$
with $SU(3)$, $SU(4)$ and $SU(5)$ spectral covers, respectively.
Assuming the minimal ${\cal Z}_2$ monodromy, we determine the homology classes and the associated spectra after flux breaking for each case. Our analysis includes
the GUT singlets which have hitherto been ignored but which
play a crucial role in phenomenology.
Using these results we construct an $E_{6}$ based model that demonstrates, for the first time, that it is possible to construct a phenomenologically viable
model which leads to the MSSM at low energies. The exotics that result from flux breaking all get a large mass when singlet fields acquire vacuum expectation values driven by D- and F-flatness. Due to the  underlying GUT symmetry and the $U(1)$s descending from $E_{8}$, bare baryon- and lepton-number violating terms are forbidden up to  and including dimension 5. As a result nucleon decay is naturally suppressed below present bounds. The $\mu$-term is forbidden by the $U(1)$ but is generated at the SUSY breaking scale when a further singlet field acquires a TeV scale vacuum expectation value, driven by the spontaneous breaking of the electroweak symmetry. After including the effect of flux and instanton corrections acceptable quark and charged lepton masses and mixing angles can be obtained. Neutrinos get a mass from the see-saw mechanism through their coupling to singlet neutrinos that acquire large Majorana mass as a result of the monodromy.
 \end{abstract}

 \end{titlepage}

\thispagestyle{empty}
\vfill
\newpage

\setcounter{page}{1}

\section{Introduction}
Almost forty years after their inception, Grand Unified Theories (GUTs)
\cite{Georgi:1974sy} remain a tantalising combination of successes and challenges
that still provides our best glimpse into the possible unity of all particle forces.
By combining GUTs with supersymmetry (SUSY) the possibility of gauge coupling unification at a higher scale remains an attractive possibility, while the hierarchy of scales is stabilised by the non-renormalisation theorem of SUSY, and gauge mediated proton decay is suppressed below current limits. However new dimension-5 operators can occur in SUSY models which may destabilise the proton, and R-parity violating operators (baryon- and lepton-number violating) must be very carefully controlled to avoid phenomenological disasters. In addition the SUSY Higgs/Higgsino mass parameter $\mu$ must be forbidden at leading order (the GUT or Planck scale), but then subsequently must be generated at the TeV scale. Moreover SUSY GUTs do not explain why there are three chiral families of quarks and leptons in complete $SU(5)$ representations, with one pair of Higgs doublets,  $H_u$ and $H_d$, in incomplete $SU(5)$ representations
(i.e. without their colour triplet $SU(5)$ partners). Nor do they shed much light on the origin of the pattern of Yukawa couplings, although $b-\tau$ unification predicted by $SU(5)$ remains viable.

Recently there has been considerable activity~\cite{Donagi:2008ca,Beasley:2008dc,Donagi:2008kj,Beasley:2008kw,Blumenhagen:2009yv,Hayashi:2008ba}
 in the reformulation of GUTs as 8D theories arising from F-theory versions of string theory~\cite{Vafa:1996xn} (for reviews and related work see e.g. \cite{Denef:2008wq,Weigand:2010wm,Heckman:2010bq,Grimm:2010ks}).
The reason for the renewed interest is that F-theory provides new opportunities
for addressing some of the above outstanding issues facing GUTs, such as GUT breaking
and Higgs doublet-triplet splitting by flux~\cite{Beasley:2008kw,Donagi:2008kj}. The original formulation of such theories on a del Pezzo surface~\cite{Beasley:2008dc} allows for gravity to be decoupled, $M_{GUT}\ll M_{Planck}$, simplifying the analysis of possible effective GUT models. In this the GUT gauge group lives on seven branes wrapping the del Pezzo surface, while quarks and
leptons and Higgs live on restricted (complex) matter curves constituting the intersections of
 this surface with other seven branes. Yukawa interactions occur at points
on the  surface at the intersection of three matter curves.

Using this structure there has been  remarkable progress on model building in F-theory over
 the last two  or three years~\cite{Heckman:2008qa}-\cite{Grimm:2011tb}. A considerable
 amount  of this work deals with the  reconciliation of F-theory models
  with the low energy Standard Model and the related phenomenology. These
  include papers related to fermion mass structure and the computation of  Yukawa
  couplings in the context of F-theory and del Pezzo singularities~\cite{Heckman:2008qa,Font:2008id,Conlon:2009qq,Dudas:2009hu,King:2010mq,Dudas:2010zb,Leontaris:2010zd,Cecotti:2009zf,Ludeling:2011en,Krippendorf:2010hj,Aparicio:2011jx}.
  In particular, some interesting mechanisms were suggested to generate Yukawa hierarchy either
  with the use of fluxes~\cite{Heckman:2008qa,Cecotti:2009zf}  and the notion of T-branes~\cite{Cecotti:2010bp}  or with the implementation of the Froggatt-Nielsen mechanism~\cite{Dudas:2009hu,King:2010mq,Dudas:2010zb,Leontaris:2010zd,Ludeling:2011en}.
More specifically, in~\cite{Cecotti:2009zf} it is argued that when three-form fluxes are turned on
in F-theory compactifications, rank-one fermion mass matrices are modified, leading to
masses for lighter generations and CKM mixing.
Ibanez et al~\cite{Aparicio:2011jx} have recently shown that flux and instanton effects can generate a realistic hierarchy of fermion masses. In the F-theory context, such non-perturbative contributions were computed in~\cite{Marchesano:2009rz},
although the magnitude of such corrections remains somewhat unclear.

Larger GUT groups than $SU(5)$ have also been considered, such as the
F-theory $E_6$ model of ref \cite{Chen:2010tg} where non-Abelian fluxes are introduced to break the
symmetry.
Flipped $SU(5)$~\cite{Heckman:2008qa,Jiang:2009za,King:2010mq, Kuflik:2010dg,Chen:2010tp}
has also been considered,
including an attempt using an $SU(4)$ spectral cover~\cite{Chung:2010bn}.
Some examples of $SO(10)$ F-theory models were also considered in~\cite{Heckman:2008qa,Chen:2009me,Chen:2010ts}.

Many (or all) of these models predict exotic states below the unification scale,
and the renormalization group (RG) analysis of gauge coupling unification including the effect of such states and flux effects has been discussed in a series of papers
\cite{Blumenhagen:2008aw}-\cite{Leontaris:2011tw}. Other phenomenological issues such as
neutrinos from KK-modes\cite{Bouchard:2009bu},
proton decay \cite{Grimm:2010ez} and the origin of CP violation~\cite{Heckman:2009de}
have also been discussed.
The possibility of obtaining the Standard Model directly from F-theory \cite{Choi:2010nf}
has also been considered.

Following this work some generic challenges have been identified that result from  the
highly constrained nature of the constructions, in particular the constraints related to the compatibility of
unification (due to the appearance of exotics),  the suppression of proton decay (due to R-parity violating operators and dimension-5 operators) the suppression of the $\mu$ term and the generation of realistic Yukawa couplings. These occur
when flux is used to break
the GUT group and generate doublet-triplet splitting. To date no fully realistic model has been constructed using just the symmetries descending from the underlying unified gauge group \cite{Dudas:2009hu,Ludeling:2011en,Dolan:2011iu} and this provides additional
motivation for the present paper.

In this paper we classify  semi-local F-theory GUTs arising from
a single $E_8$ point of local enhancement, leading to simple GUT groups
based on $E_6$, $SO(10)$ and $SU(5)$ on the del Pezzo surface
with $SU(3)$, $SU(4)$ and $SU(5)$ spectral covers, respectively.
In the semi-local approach to F-theory, it is normally assumed that there exists a single point of
$E_8$ enhancement in the internal geometry~\cite{Heckman:2009mn}, from which all the interactions come.  We study the matter that descends from the adjoint of $E_8$ for the following breaking patterns:
\bea
 E_8 & \supset &  E_6 \times SU(3)_{\perp} \nonumber\\
 E_8 & \supset & SO(10) \times SU(4)_{\perp} \nonumber\\
E_8 & \supset & SU(5) \times SU(5)_{\perp}\nonumber
\eea
Assuming the minimal case of a  ${\cal Z}_2$  monodromy, we discuss the flux breaking
and homology classes of the spectrum for each case,
and provide a dictionary relating the representations of
the different GUT groups that can lead to new physical insights into
model building.
We assume that all breaking of the GUT gauge group to the Standard Model occurs when fluxes associated with the $U(1)$s in the perpendicular groups are turned on.  To determine the chiral spectrum we need to know how the fluxes restrict on the various matter curves.  There two kinds of flux that we need to consider.  Firstly, we have the fluxes associated with the  $U(1)$s remaining after the imposition of a  ${\cal Z}_2$  monodromy and the perpendicular gauge group has been broken.  These fluxes determine the chirality of the  complete GUT representations.  Secondly, we have the {hypercharge} flux inside the GUT group, which breaks the remaining gauge symmetry down to that of the Standard Model.  To determine the effect of the flux it is convenient to work in the spectral cover approach. Using this we determine the spectrum after flux breaking. Our analysis includes
the singlet spectrum which has hitherto been ignored but which
plays a crucial role in phenomenology.

{As an example of an application of our results we consider the construction of a viable low-energy-model in which the $U(1)$ symmetries and flux effects answer all the challenges posed above. We start with the identification of R-parity in an $E_{6}$ GUT. After flux breaking the model has some undesirable features but it proves possible to eliminate these by relaxing the $E_{6}$ constraints on the spectrum. However the dangerous $R-$parity violating operators are still forbidden. In addition the dimension 5 nucleon decay operators, allowed by $R$-parity, are also forbidden due to the $U(1)$ global symmetries of the model.
\newline\indent
Due to the flux breaking the spectrum has additional vector-like states beyond those of the minimal supersymmetric extension of the Standard Model (MSSM). We show that these exotic states get a large mass, close to the compactification scale, if certain SM (and $SU(5)$) singlet fields acquire vacuum expectation values (VEV). We identify the necessary singlet fields and show that these VEVs are needed for D- and F-flatness of the scalar potential, the VEVs being driven close to the compactification scale. Moreover we show that these VEVs do not re-introduce terms that can give rapid proton decay. In addition we show that a $\mu$ term of the SUSY breaking scale is naturally obtained through another singlet field acquiring a TeV scale VEV.
\newline\indent
Finally we show that the model may have a realistic structure for the quark and charged lepton masses in which the light generation masses and mixings are driven by flux and instanton effects. The neutrinos can get mass from the (type I) ``see-saw'' through the coupling of the doublet neutrinos to singlet neutrinos that acquire Majorana mass due to the monodromy.}

{The layout of the remainder of the paper is as follows.
In Section 2 we discuss a dictionary connecting the $SU(3)_{\perp}\times E_6$
and $SU(5)_{\perp}\times SU(5)$ {representations} and their $U(1)_{\perp}$ charges. This proves to be useful when
constructing viable models. In Section 3, using the spectral cover approach, we determine the homology of the gauge non-singlet and gauge singlet fields for the three breaking patterns given above. We discuss how the homology gives constraints on the spectrum after flux breaking. In Section 4 we discuss the D-and F-flatness conditions that apply for the case only the $SU(5)$ singlets acquire VEVs.
In Section 5 we discuss the construction of a realistic model that, after a definite set of singlet VEVs are switched on, has just the MSSM spectrum and renormalisable couplings. Using the results of Section 4 we show that the F-and D-flatness conditions do indeed induce the needed VEVs and we determine their relative magnitude. We show that, due to the $U(1)$ symmetries and the underlying GUT structure, the model avoids dangerous baryon- and lepton-number terms up to and including the dangerous dimension 5 operators. We also discuss how a $\mu$ term of the required order is generated. Finally we consider the structure of the quark, charged lepton and neutrino masses and mixings and show that they can be realistic.}


\section{Group Theory Dictionary Between $E_6$ and $SU(5)$  \label{Group}}

In this paper we are concerned with the sequence of rank preserving symmetry breakings, induced by flux breaking. Starting from the highest allowed symmetry in elliptic fibration, that is the $E_8$
exceptional group, there exists a variety of breaking patterns to obtain the
Standard Model gauge symmetry. A complete classification of these possibilities
from the F-theory perspective has been given in the appendix of ref~\cite{Beasley:2008kw}.
 Here, we shall be interested in the general embeddings discussed in the Introduction, where the adjoint of $E_8$ decomposes in each case as
\bea
 E_8 & \supset &  E_6 \times SU(3)_{\perp} \label{SU(3)} \label{E61}\\
248 & \rightarrow &  (78,1)+(27,3)+(\overline{27},\overline{3})+(1,8) \label{E6000} \\
 E_8 & \supset & SO(10) \times SU(4)_{\perp} \label{SU(4)}\\
 248 & \rightarrow & (1,15)+(45,1)+(10,6)+(16,4)+(\ov{16},\ov{4}) \\
E_8 & \supset & SU(5) \times SU(5)_{\perp} \label{SU(5)} \\
 248 & \rightarrow & (24,1)+(1,24) +(10,5)+(\overline{5},10)+(\overline{5}, \overline{10})+(5,\overline{10})
 \label{DP00}.
\eea
 The last one in particular has been extensively studied by many
 authors including~\cite{Beasley:2008kw,Donagi:2009ra,
Marsano:2009gv,Dudas:2009hu,King:2010mq}.
In this case, the $SU(5)_{GUT}$ is the maximal subgroup $SU(5)\in E_8$
while the corresponding matter content transforms non-trivially under the Cartan subalgebra
of $SU(5)_{\perp}$ with weight vectors $t_{1,...,5}$ satisfying
\be
t_1+t_2+t_3+t_4+t_5=0.
\ee
In principle, the superpotential can be maximally constrained by four $U(1)$'s according
to the breaking pattern
\be
E_8  \supset SU(5)\times SU(5)_\perp \rightarrow SU(5)\times U(1)_\perp^4
\label{DP0}
\ee
The $5$ representation of $SU(5)_\perp$ may be expressed in the conventional basis of the five weight vectors $t_i$
in which the 4 Cartan generators corresponding to  $U(1)_\perp^4$ are expressed as:
\bea
&& H_1=\frac{1}{2}{\rm diag}(1,-1,0,0,0), \ \  H_2=\frac{1}{2\sqrt{3}}{\rm diag}(1,1,-2,0,0),\nonumber \\
&& H_3=\frac{1}{2\sqrt{6}}{\rm diag}(1,1,1,-3,0),
\ \ H_4=\frac{1}{2\sqrt{10}}{\rm diag}(1,1,1,1,-4).
\label{Hi}
\eea
In general, however, there is an action on $t_i$'s of a non-trivial monodromy group which is
a subgroup of the Weyl group $W(SU(5)_{\perp})=S_5$. Such subgroups are the alternating
groups ${\cal A}_n$, the dihedral groups ${\cal D}_n$ and cyclic groups ${\cal Z}_n$, $n\le 5$ and the
Klein four-group ${\cal Z}_2\times {\cal Z}_2$.
Throughout this paper we shall assume the minimal  ${\cal Z}_2$ monodromy, $t_1\leftrightarrow t_2$.

It is of interest to consider the possibility of a sequence of flux breaking, which may be associated with different scales. Here we consider the sequence

\bea
E_8
& \rightarrow & {E_6}\times U(1)_\perp^2 \label{E6} \label{E60}\\
& \rightarrow & SO(10)\times U(1)_{\psi}\times U(1)_\perp^2 \label{SO(10)} \\
& \rightarrow & SU(5)\times  U(1)_{\chi} \times U(1)_{\psi} \times  U(1)_\perp^2 . \label{SU(5)0}
\eea
which for the $E_{6}$ representations gives
\bea
78 &\rightarrow & [24_{(0,0)} + 10_{(4,0)} + \overline{10}_{(-4,0)} + 1_{(0,0)}]_{45}\nonumber\\
&+& [10_{(-1,-3)} + \overline{5}_{(3,-3)} + 1_{(-5,-3)}]_{16}\nonumber\\
&+& [\overline{10}_{(1,3)} + {5}_{(-3,3)} + 1_{(5,3)}]_{\overline{16}}\nonumber\\
&+& [1_{(0,0)}]_{1}\nonumber\\
27 &\rightarrow &  [10_{(-1,1)} + \overline{5}_{(3,1)} + 1_{(-5,1)}]_{16}\nonumber\\
&+& [ {5}_{(2,-2)} + \overline{5}_{(-2,-2)}]_{\overline{10}}\nonumber\\
&+& [1_{(0,4)}]_{1} \label{E601}
\eea
where the subscripts refer to the $U(1)_{\chi},U(1)_{\psi}$ charges and $SO(10)$ representation respectively.  It is convenient to choose a basis for the weight vectors such that the charge generators have the form
\bea
Q_{\chi}&\propto &{\rm diag}[-1,-1,-1,-1,4]\nonumber \\
Q_{\psi}&\propto &{\rm diag}[1,1,1,-3,0]  \nonumber \\
Q_{\perp}&\propto &{\rm diag}[1,1,-2,0,0]
\label{charges}
\eea
\begin{table}[t]
\begin{center}
\small
\begin{tabular}{|c|c|c|c|}
\hline
$E_6$ & $SO(10)$ & $SU(5)$  & Weight vector \\
\hline
$27_{t_1'}$ & $16$ & $\overline{5}_3$ & $t_1+t_5$ \\
\hline
$27_{t_1'}$ & $16$ & $10_M$ & $t_1$ \\
\hline
$27_{t_1'}$ & $16$ & $\theta_{15}$ & $t_1-t_5$ \\
\hline
$27_{t_1'}$ & $10$ & $5_1$ & $-t_1-t_3$ \\
\hline
$27_{t_1'}$ & $10$ & $\overline{5}_2$ & $t_1+t_4$\\
\hline
$27_{t_1'}$ & $1$ & $\theta_{14}$ & $t_1-t_4$ \\
\hline
$27_{t_3'}$ & $16$ & $\overline{5}_5$ & $t_3+t_5$ \\
\hline
$27_{t_3'}$ & $16$ & $10_2$ & $t_3$ \\
\hline
$27_{t_3'}$ & $16$ & $\theta_{35}$ & $t_3-t_5$ \\
\hline
$27_{t_3'}$ & $10$ & $5_{H_u}$ & $-2t_1$ \\
\hline
$27_{t_3'}$ & $10$ & $\overline{5}_4$ & $t_3+t_4$ \\
\hline
$27_{t_3'}$ & $1$ & $\theta_{34}$ & $t_3-t_4$ \\
\hline
\end{tabular}
\end{center}
\caption{\small Complete $27$s of $E_6$ and their $SO(10)$ and $SU(5)$ decompositions.
For the $SU(5)$ states we use the  notation of ref~\cite{Dudas:2009hu} where indices
in $5_i, 10_j$ representations are associated to the corresponding matter curves
$\Sigma_{5_i}, \Sigma_{10_j}$.}
\label{t1}
\end{table}%
where $Q_{\perp}$ is the charge of the $U(1)_{\perp}$ in the breaking pattern of Eq. (\ref{E60}) that remains after imposing the $t_1\leftrightarrow t_2$ monodromy.
This is in fact the same as the conventional basis for the  $SU(5)_\perp$ generators in Eq. (\ref{Hi}),
and the normalisation of the generators is given by identifying,
\be
H_1=H_1', \ \ H_2=Q_{\perp}, \ \ H_3 = Q_{\psi}, \ \ H_4 = -Q_{\chi}.
\ee
This almost trivial equivalence shows that the $SU(5)_{GUT}$ states in Eq. (\ref{DP00})
have well defined $E_6$ charges $Q_{\chi}$ and $Q_{\psi}$.
For example $SU(5)$ singlets will in general carry $Q_{\chi}$ and $Q_{\psi}$ charges
which originate from $E_6$ and which may be unbroken. The equivalence will provide
insights into both anomaly cancellation and the origin of R-parity for example,
in terms of the underlying $E_6$ structure, in the explicit models discussed later.
Throughout this paper we shall assume the minimal  ${\cal Z}_2$ monodromy, $t_1\leftrightarrow t_2$ \cite{Dudas:2010zb}
which trivially corresponds to the minimal ${\cal Z}_2$ monodromy, $t'_1\leftrightarrow t'_2$.
It is clear from Eq. (\ref{Hi}) that this corresponds to $H_1=H_1'$ being broken leaving only three independent Cartan symmetries $\{ H_2, H_3, H_4\}$ or equivalently $\{Q_{\perp}, Q_{\psi}, Q_{\chi}\}$.

In this basis the weight vectors $t_{1}',t_{2}',t_{3}'$ ($t_{1}'+t_{2}'+t_{3}'=0$) of $SU(3)_{\perp}$ are related to the $SU(5)_{\perp}$ weight vectors by $t_{i}'=t_{i}+(t_{4}+t_{5})/3,\;\;i=1,2,3$. As an example of the use of this dictionary that will play an important role when building a realistic theory we can now connect the two independent representations $27_{t_{1}'}$ and $27_{t_{3}'}$ that appear in the $E_{6}$ breaking pattern of Eq. (\ref{E60}) to the $SU(5)$ representations of Eq. (\ref{SU(5)0}). These are shown in Table \ref{t1} with $SU(5)$ states given in the notation of \cite{Dudas:2009hu}.

\section{Flux breaking and matter content in F-theory GUTs}
In this Section we determine the light matter content that results if the underlying $E_{8}$ GUT group is broken by flux. We are interested in the cases that the unbroken gauge group is $E_{6},\; SO(10)$ or $SU(5)$. One reason to study these patterns of breaking is because subsequent breaking to the Standard Model may proceed via the normal Higgs mechanism with fields acquiring VEVs along flat directions. In this knowledge of the multiplet content before such breaking is crucial. A second reason to study these patterns is because it can suggest promising phenomenological models based on a high degree of unification even though they are subsequently further broken by flux to just the Standard Model. We will present a viable model in the next Section.

We proceed by studying the spectral cover of the transverse groups for the three cases of interest $E_{6}\times SU(3)_{\perp},\;SO(10)\times SU(4)_{\perp}$ and $SU(5)\times SU(5)_{\perp}$. This will allow us to determine the homology of the matter fields and hence the effect of flux breaking. A novel feature of this work is that we determine for the first time the full singlet homology.

 \subsection{$SU(3)_{\perp}$ Spectral Cover}

$E_6$ models are quite attractive and have been extensively studied
in compactifications on Calabi-Yau manifolds, in the context of
 the heterotic superstring with underlying $E_8 \times E_8$ symmetry (see
\cite{Greene:1986bm,Greene:1986jb,Greene:1987xh} and references therein). Furthermore,
recent phenomenological investigations based on string motivated
versions with  $E_6$   gauge symmetry have inspired the exceptional supersymmetric
 standard model~\cite{King:2005jy}. This is distinguished from the minimal
 one by the appearance of an additional $Z'$  boson
 and extra matter content at the TeV scale.  Interestingly, although these new ingredients are also potentially present
 in the F-theory $E_6$-analogue, they are subject to constraints from flux restrictions
on matter curves and the topological properties of the compact manifold, and in the model considered later the $Z'$  boson has a GUT scale mass and the extra matter also has a similarly large mass.

In the context of F-theory in which the GUT group on the brane is $E_6$,
 we need to look at the breaking
\begin{equation}
 E_8  \rightarrow E_6 \times SU(3)_{\perp}
\end{equation}

\noindent We can determine what matter curves arise by decomposing the adjoint of $E_8$ as follows

\begin{equation}
 248  \rightarrow  (78,1)+(27,3)+(\overline{27},\overline{3})+(1,8)
\end{equation}

\noindent The $E_{6}$ content consists of three 27s (and $\ov{27}s$) plus eight singlet matter curves.  In terms of the weight vectors $t_i$, i=1,2,3, of $SU(3)_{\perp}$ the equations of these curves are
\begin{align}
 \Sigma_{27} &: t_i=0 \label{27 curve} \\
\Sigma_{1} &: \pm(t_i-t_j)=0 \ \ i \neq j \label{1 curve SU(3)}
\end{align}
The $SU(3)$ spectral cover is found by determining two sections $U,V$ of the
projective bundle ${\cal P}({\cal O}\oplus K_S)$ over the compact surface $S$.
Let $c_1$ be the 1$^{st}$ Chern class of the {\it tangent} bundle to $S_{GUT}$ and $c_1(NS)=-t$ that
of the {\it normal} bundle. Note that
the homological classes are $[U]=-t$ and $[V]=c_1-t$. The $SU(3)$ spectral cover is
\ba
{\cal C}^{3}:&&b_0U^3+b_2UV^2+b_3V^3=0\label{C3}
\ea
 Associated to this is the polynomial
\ba
P_3=\sum_{k=0}^3 b_ks^{3-k}\;=\;b_3+b_2s+b_1s^2+b_0s^3\label{P3}
\ea
where we have introduced the affine parameter $s=\frac{U}{V}$.

\noindent We define for convenience  $\eta = 6c_1-t$ and, as usual, we demand that the coefficients $b_k$ are sections of
\ba
b_k:&&[b_k]= \eta-k\,c_1\label{bhc}
\ea
\noindent where $k$ spans the integers $k=1,2,3,4,5$.
 The roots of the spectral cover equation
\[0=b_3+b_2s+b_0s^3\propto \prod_{i=1}^3(s+t'_i)\]
are identified as the $SU(3)_{\perp}$ weight vectors  $t'_i$.
In the above the coefficient $b_1$ is taken to be zero since it corresponds to the sum
of the roots which, for $SU(n)$, is always zero, $\sum_i t'_i=0$.
\subsubsection{$27$ and $\overline{27}$ fields}
The coefficient $b_3$ is equal to the product of
the roots, i.e. $b_3=t_1't_2't_3'$ and the  $\Sigma_{27}$ curves where
 the corresponding matter  multiplets are  localized are determined by
 its three zeros
\ba
\Sigma_{27_i},\;\;\; b_3=\prod_{i=1}^3t'_i=0\ra t'_i=0,\;\;\; i=1,2,3\label{27s}
\ea
To obtain different curves for $27$'s we need to split the spectral cover. (If the polynomial
is not factorized, there is only one matter curve). There are two possible
ways to split a  third degree polynomial:  either to a binomial-monomial ($2-1$) or
to three monomials ($1-1-1$). Since we need to impose
a monodromy action, we choose this to be ${\cal Z}_2$ and therefore we get a ($2-1$) split.
The  ${\cal Z}_2$ monodromy corresponds to the
following split of the spectral cover equation
\ba
0=\Pi_3(s)&=&(a_1+a_2s+a_3s^2)(a_4+a_5s)\nn\\
          &=&a_1a_4+(a_2a_4+a_1a_5)s+(a_2a_5+a_3a_4)s^2+a_3a_5s^3\label{P3fact}
\ea
with $s=U/V$ and $a_i$  coefficients, constituting sections of line bundles
each of them being of specific  Chern class to be determined.

 The first bracket contains the polynomial factor that corresponds to the  ${\cal Z}_2$
monodromy  $t'_1\leftrightarrow t'_2$, so that the corresponding
 two $\Sigma_{27}$ curves lift to a common  one in the spectral cover.
The $\Sigma_{27}$ curves are found setting $s=0$ in the polynomial
\ba
b_3\equiv\Pi_3(0)=a_1a_4=0&\ra& a_1=0, \, a_4=0\nn
\ea
Thus, after the monodromy action, we obtain two matter curves. When building a realistic theory it is necessary to assign the three families of quarks and leptons and the Higgs to these curves. As there are more than one way to do this, the optimal choice will be dictated by phenomenology.

To determine the distribution of families and Higgs on the two matter curves we need
to know how the flux restricts on the available curves. To do this we first determine their homology
classes  $[a_{k}]$ corresponding to the sections $a_k,\;k=1,4$.  This can be done
comparing the coefficients of  Eqs. (\ref{P3},\ref{P3fact}). We get
\ba
b_0&=&a_3a_5\nn\\
b_1&=&a_2a_5+a_3a_4=0\nn\\
b_2&=&a_2a_4+a_1a_5\nn\\
b_3&=&a_1a_4\nn
\ea
The homology classes $[b_k]$ of the sections $b_k$ are given in Eq. (\ref{bhc}),
 while those of $a_i$ can be determined
by the system of linear  equations in one to one correspondence  with
the above relations. This linear system consists of four equations with five unknowns $[a_i]$,
 therefore we can  solve the system
in terms of  one arbitrary parameter.
Let $a_5$ be of some unspecified homology class $[a_5]=\chi $. For the remaining $a_i$,
we find that they are sections of
\ba
[a_1]=\eta-2 c_1-\chi,[a_2]=\eta-\chi-c_1,\;[a_3]=\eta-\chi,\;[a_4]=\chi-c_1,\;[a_5]=\chi
\ea
For the two curves we obtain the results of Table \ref{e6homology}.
\begin{table}
\begin{center}
\begin{tabular}{lcl}
Matter& Section& Homology \\
\hline
$27_{t_{1,2}}$ &  $a_1$ & $\eta-2c_1-\chi$ \\
$27_{t_{3}}$ &  $a_4$ & $\chi-c_1$ \\
\end{tabular}
\end{center}
\caption{\small{The three columns show  the quantum numbers of matter curves under
$E_6\times U(1)_{t_i}$, the section and  the homology class.}}
\label{e6homology}
\end{table}
For the homology classes of the two curves ${\cal C}^3={\cal C}_{t_{1,2}}{\cal C}_{t_3}$
 from Eq. (\ref{C3}) we get
\ba
{\cal C}_{t_{1,2}}&=&a_1V^2+a_2UV+a_2U^2\\
{\cal C}_{t_3}&=&a_4V+a_5U
\ea
so that their homology classes are given by
\[ [{\cal C}_{t_{1,2}}]=\eta-\chi-2\,t,\;\;[{\cal C}_{t_{3}}]=\chi-t\]

Using the data of Table  \ref{e6homology}, we can turn on a ${\cal F}_{U(1)}$ flux on the external $U(1)$
and find the restriction on the curves of $27$'s:
\ba
n_{t_1}={\cal F}_{U(1)}\cdot (\eta-\chi-2 c_1)&;&n_{t_3}= {\cal F}_{U(1)}\cdot (\chi- c_1)
\label{flux}
\ea
 These determine the chiral content  of states  arising from
  the decomposition of $27$'s along  the matter curves. We have also
 seen that ${\chi}$ is some unspecified homology class (associated to $a_5$)
  and it can be chosen at will. For acceptable choices it
 can be seen from Table  \ref{e6homology} the two curves cannot be of the same homology class.
Since the two curves belong to different homology classes, in general flux
restricts differently on them. The two conditions can be  combined as follows
 \ba
 n_{t_3}+ n_{t_1}&=& {\cal F}_{U(1)}\cdot (\eta- 3 c_1)={\cal F}_{U(1)}\cdot (3 c_1-t)
 \label{n_flux}
 \ea
 From Eq. (\ref{n_flux}) we deduce that if ${\cal F}_{U(1)}\cdot (3 c_1-t)= 0$, then
 $ n_{t_3}=- n_{t_1}$ i.e., we get opposite flux restrictions on $27_{t_1}$ and ${27}_{t_3}$.
 Notice that the choice ${\cal F}_{U(1)}\cdot c_1\ne 0$  implies that
 the corresponding gauge boson becomes massive. This is not a problem however, for
 the extra $U(1)$'s  that do not participate in the hypercharge definition~\footnote{
 For a recent work on the  $U(1)$ symmetries in F-theory see~\cite{Grimm:2011tb}.}.

\subsubsection{$E_{6}$ singlets}

Singlet fields are important for the construction of the low energy effective
field theory model. Some of them may develop VEVs  that can  be used to create mass terms for the fermion generations and make massive other potentially dangerous fields mediating proton
decay. In certain models, those carrying charges under the weights $t_i'$ undergoing a monodromy action can play the role of the right handed neutrino~\cite{Bouchard:2009bu}.
Because of their central role in phenomenology, we need to determine
their exact properties.

 An $E_{6}$ singlet $\theta_{ij}$ lies in the $t_i'-t_j'$ direction
of the corresponding Cartan subalgebra and the
 polynomial describing them  is
\begin{equation}
P = b_{0}^{3} (s + t'_1 - t'_2 )(s + t'_2 - t'_1 )(s + t'_1 - t'_3 )(s + t'_3 - t'_1 )(s + t'_2 - t'_3 )(s + t'_3 - t'_2 )
\end{equation}

\noindent where $t'_3 = -t'_1 - t'_2$.  In the usual way, we set $s=0$ to find the equations for the curves, and denote $P(s=0) \equiv P_0$.  As such, the singlet curves are given by $P_0 = 0$.
We can  express $P_0$ in terms of the $b_i$ to obtain

\begin{equation}
P_0 = -4 b_2^3 - 27 b_3^2 b_0
\end{equation}

\noindent Now in writing this in terms of the $a_i$, we solve the constraint $b_1 = 0$ with the relations $a_3 = k a_5$, $a_2 = -k a_4$.  This leads to the equation

\begin{equation}
P_0 = -(4 a_1 a_5 - k a_4^2) (a_1 a_5 + 2 k a_4^2)^2
\end{equation}

\noindent As such, the equations for the singlets are

\begin{align}
\theta_{12} & \rightarrow (4 a_1 a_5 - k a_4^2) \\
\theta_{13} & \rightarrow (a_1 a_5 + 2 k a_4^2) \\
\theta_{23} & \rightarrow (a_1 a_5 + 2 k a_4^2)
\end{align}

\noindent We can see from this that the singlets all have the same homology class

\ba
\theta_{ij}:&& [\theta_{ij}]=\eta - 2 c_1\label{singhomol}
\ea

 \subsection{$SU(4)_{\perp}$ Spectral Cover}

$SO(10)$ GUT is one of the most promising  Unified Theories and the smallest
one incorporating the right-handed neutrino into the same multiplet with
the remaining fundamental particles (quarks and leptons).
For the case that the GUT group on the brane is SO(10) we need to consider the breaking

\begin{equation}
 E_8 \rightarrow SO(10) \times SU(4)_{\perp}
\end{equation}

\noindent We can determine which matter curves arise from the decomposition of the adjoint of $E_8$:

\begin{equation}
 248 \rightarrow (1,15)+(45,1)+(10,6)+(16,4)+(\ov{16},\ov{4})
\end{equation}

\noindent Thus there are four 16 (and $\ov{16}$) matter curves, six 10 matter curves, and fifteen singlets.  The equations for these curves in terms of the weight vectors $t_i$, i=1,2,3,4, of $SU(4)_{\perp}$ are

\begin{align}
 \Sigma_{16} &: t_i=0 \label{16 curve} \\
\Sigma_{10} &: (-t_i-t_j)=0, \ \ i \neq j \label{10 curve} \\
\Sigma_{1} &: \pm(t_i-t_j)=0 \ \ i \neq j \label{1 curve SO(10)}
\end{align}

\noindent where $\sum_i t_i = 0$.  In order to determine how fluxes restrict on these matter curves, taking into account the effects of monodromy, the spectral cover approach is used.  The SU(4) spectral cover is a hypersurface given by the constraint

\ba
{\cal C}^{4}: &&b_0 U^4+b_1 V U^3 + b_2 V^2 U^2+b_3 V^3 U +b_4 V^4 = 0\label{spectral cover U V}
\ea

\noindent We can set an affine parameter $s=U/V$ which Eq. \eqref{spectral cover U V} is a polynomial in, and whose 5 roots are the $t_i$.  This $s$ can be equated with the value of the Higgs field that breaks the $E_8$ gauge theory.  In terms of s, we have:

\begin{align}
{\cal C}^{4} &= b_0 s^4 + b_1 s^3 + b_2 s^2 + b_3 s + b_4 = 0 \label{16 poly b} \\
&=b_0 (s+t_1)(s+t_2)(s+t_3)(s+t_4) = 0 \label{16 poly t}
\end{align}

\noindent where the second line reflects the fact that the $t_i$ are the roots of the polynomial.  This polynomial describes the 16 matter curves, which are given by setting $s$ to zero in the above equations, leading to $b_4 = 0$.  Equations for the $b$'s in terms of the $t$'s can be found
by comparing powers of $s$ in Eqs. \eqref{16 poly b} and \eqref{16 poly t}.
 This leads to the following equations, once $t_4$ has been eliminated by using the fact that the sum of the $t_i$ is zero:

\begin{align}
b_1 &= -b_0 (t_1 +t_2 +t_3 + t_4) = 0 \label{b1} \\
b_2 &= b_0 (t_1^2 +t_2^2 +t_3^2 +t_1 t_2 + t_2 t_3 +t_1 t_3) \label{b2} \\
b_3 &= b_0 (t_1 + t_2)(t_2 + t_3)(t_1 + t_3) \label{b3} \\
b_4 &= -b_0 t_1 t_2 t_3 (t_1 + t_2 + t_3) \label{b4}
\end{align}

\noindent It can be seen that the equation $b_4=0$ does indeed reproduce Eq. \eqref{16 curve} for the 16 matter curves in terms of the $t_i$.

\subsubsection{ ${\cal Z}_2$  Monodromy}

Imposing a  ${\cal Z}_2$  monodromy
implies the splitting of Eq. \eqref{16 poly b} as follows

\begin{equation}
{\cal C}^{4} = (a_1 + a_2 s + a_3 s^2)(a_4 + a_5 s)(a_6 + a_7 s) \label{Z2}
\end{equation}

\noindent The first bracket is quadratic in $s$ reflecting the fact that we have chosen a ${\cal Z}_2$  monodromy, which in the weight language corresponds to an identification of two weights $t_1 \leftrightarrow t_2$.  We can now match powers of $s$ in Eqs. \eqref{16 poly b} and \eqref{Z2} to get equations for the $b_i$ in terms of the $a_i$.
\begin{align}
b_0 &= a_3 a_5 a_7 \label{b0 a} \\
b_1 &= a_2 a_5 a_7 + a_3 a_5 a_6 + a_3 a_4 a_7 \label{b1 a} \\
b_2 &= a_1 a_5 a_7 + a_2 a_4 a_7 + a_2 a_5 a_6 + a_3 a_4 a_6 \label{b2 a} \\
b_3 &= a_1 a_4 a_7 + a_1 a_5 a_6 + a_2 a_4 a_6 \label{b3 a} \\
b_4 &= a_1 a_4 a_6 \label{b4 a}
\end{align}

\noindent Solving for $b_1 = 0$ gives
\begin{align}
a_2 &= -\gamma (a_5 a_6 +a_4 a_7) \label{a2 solve} \\
a_3 & = \gamma a_5 a_7 \label{a3 solve}
\end{align}

\noindent where $\gamma$ is unspecified.  Now we can demand that the homology classes of the $b_n$ are

\begin{equation}
[b_n]=\eta-n c_1 \label{b homology}
\end{equation}

\noindent where, as before, $\eta=6c_1 -t$, $c_1$ is the first Chern class of the tangent bundle to $S_{GUT}$ and $-t$ is the first Chern class of the normal bundle.  We can now determine the homology classes of the $a_i$ coefficients by using Eqs. \eqref{b0 a}-\eqref{b4 a}, setting the homology class of a given $b_n$ equal to the homology class of each product of $a_i$s on the left hand side of the appropriate equation.  This leads to
\begin{align}
\eta &= [a_3] + [a_5] + [a_7] \label{sim 1} \\
\eta-c_1 &= [a_2] + [a_5] + [a_7] \label{sim 2} \\
\eta-2 c_1 &= [a_1] + [a_5] + [a_7] \label{sim 3} \\
\eta-3 c_1 &= [a_1] + [a_4] + [a_7] \label{sim 4} \\
\eta-4 c_1 &= [a_1] + [a_4] + [a_6] \label{sim 5}
\end{align}

As such, we have 5 equations in 7 unknowns, and so we can solve the equations in terms of two free parameters, which we can set as

\begin{align}
[a_5] &= \chi_5 \label{chi5} \\
[a_7] &= \chi_7 \label{chi7} \\
\tilde{\chi} &= \chi_5 + \chi_7 \label{chi tilde}
\end{align}

\noindent Solving the system of equations gives the homology classes of the remaining $a_i$

\begin{align}
[a_1] &= \eta - 2 c_1 - \tilde{\chi} \label{a1 homology} \\
[a_2] &= \eta - c_1 - \tilde{\chi} \label{a2 homology} \\
[a_3] &= \eta - \tilde{\chi} \label{a3 homology} \\
[a_4] &= -c_1 + \chi_5 \label{a4 homology} \\
[a_6] &= -c_1 + \chi_7 \label{a6 homology}
\end{align}

\noindent We now have determined the homology classes of all the $a_i$ coefficients (which are summarised in Table \ref{atable}), and can use them in order to find the homology classes of the matter curves.

\begin{table}
\begin{center}
\begin{tabular}{lcl}
Coefficient & Homology \\
\hline
$a_1$ & $\eta - 2 c_1 - \tilde{\chi}$ \\
$a_2$ & $\eta - c_1 - \tilde{\chi}$ \\
$a_3$ & $\eta - \tilde{\chi}$ \\
$a_4$ & $-c_1 + \chi_5$ \\
$a_5$ & $\chi_5$ \\
$a_6$ & $-c_1 + \chi_7$ \\
$a_7$ & $\chi_7$
\end{tabular}
\end{center}
\caption{\small{Homology classes of the $a_i$ coefficients}}
\label{atable}
\end{table}

\subsubsection{Homology of the  16 Matter Curves}

As discussed after Eq. \eqref{16 poly t}, the 16 matter curves are given by $b_4 = 0$.  From Eq. \eqref{b4 a}, this means that the equations of the 16s are

\begin{equation}
a_1=0, \ \ a_4=0, \ \ a_6=0 \label{equations 16 a}
\end{equation}

\noindent and so the homology classes are

\begin{align}
[16_{1}] &= \eta - 2 c_1 - \tilde{\chi} \label{16_1 homology} \\
[16_{2}] &= -c_1 + \chi_5 \label{16_2 homology}\\
[16_{3}] &= -c_1 + \chi_7 \label{16_3 homology}
\end{align}

\subsubsection{Homology of the 10 Matter Curves}

Just as the correct polynomial to describe the 16 matter curves was the spectral cover polynomial, the polynomial for the 10s is given by

\begin{align}
P_{10} &= b_0^2 \prod_{i<j} (s+t_i + t_j)\nonumber \\
&= b_0^2 (s-t_1 -t_2)(s+t_1 +t_2)(s-t_1 -t_3)(s+t_1 +t_3)(s-t_2 -t_3)(s+t_2 + t_3) \label{P10 factored} \\
&= s^6 + c_1 s^5 + c_2 s^4 + c_3 s^3 +c_4 s^2 + c_5 s + c_6 \label{P10 expanded}
\end{align}

\noindent where in Eq. \eqref{P10 factored}, $t_4$ has been eliminated by using $\sum_i t_i = 0$.  Comparing coefficients of $s$ between Eqs. \eqref{P10 factored} and \eqref{P10 expanded} the following equations for the $c_i$ in terms of the $t_i$
are obtained
\begin{align}
c_1 &= 0 \label{c1} \\
c_2 &= -2(t_1^2 + t_2^2 + t_3^2 + t_1 t_2 + t_1 t_3 + t_2 t_3) b_0^2 \label{c2} \\
c_3 &= 0 \label{c3} \\
c_4 &= [t_1^4 +2 t_1^3 (t_2 + t_3) + (t_2^2 + t_2 t_3 + t_3^2)^2 + t_1^2(3 t_2^2 + 8 t_2 t_3 + 3 t_3^2) \notag \\
&+ 2t_1 (t_2^3 + 4t_2^2 t_3 +4t_2 t_3^2 + t_3^3)] b_0^2 \label{c4} \\
c_5 &= 0 \label{c5} \\
c_6 &= -(t_1 + t_2)^2 (t_1 + t_3)^2 (t_2 + t_3)^2 b_0^2 \label{c6}
\end{align}

\noindent We can now use Eqs. \eqref{b1}-\eqref{b4} to write the $c_i$ coefficients in terms of the $b_i$.   The results are

\begin{align}
c_2 &= -2 b_0 b_2 \label{c2 b} \\
c_4 &= b_2^2 - 4 b_4 b_0 \label{c4 b} \\
c_6 &= -b_3^2 \label{c6 b}
\end{align}

\noindent Substituting into Eq. \eqref{P10 expanded} gives

\begin{equation}
P_{10} = s^6 -2b_0 b_2 s^4 + (b_2^2-4 b_4 b_0) s^2 -b_3^2 \label{P10 b}
\end{equation}

\noindent As in the case of the 16 polynomial, the 10 matter curves are found by setting $s$ to zero in this equation, giving $b_3^2 = 0$.  In order to know the equations and homology classes for the 10 matter curves when the monodromy is imposed, we must express this equation in terms of the $a_i$ coefficients.  From Eq. \eqref{b3 a}, we know $b_3$ in terms of the $a_i$.  Substituting Eq. \eqref{a2 solve} in for $a_2$ leads to

\begin{equation}
b_3 = (a_5 a_6 + a_4 a_7)(a_1-\gamma a_4 a_6) \label{b3 a2}
\end{equation}

\noindent As such, the 10 matter curves are defined by the equation

\begin{equation}
(a_5 a_6 + a_4 a_7)(a_1-\gamma a_4 a_6)(a_5 a_6 + a_4 a_7)(a_1-\gamma a_4 a_6)=0 \label{10 curves poly}
\end{equation}

\noindent We therefore have four 10 matter curves, two of which have homology class $[a_1] = \eta - 2 c_1 - \tilde{\chi}$, and two of which have homology class $[a_5 a_6] = [a_5]+[a_6] = -c_1 + \tilde{\chi}$.  The information about the homology classes of all the 16 and 10 matter curves is summarised in Table \ref{SU4table}.  For convenience, the following notation is introduced

\begin{align}
M &= \mathcal{F}_1 \cdot (\eta - 3c_1) \\
P &= \mathcal{F}_1 \cdot (\chi - c_1) \\
P_n &= \mathcal{F}_1 \cdot (\chi_n - c_1) \\
C &= \mathcal{F}_1 \cdot (-c_1)
\end{align}

\begin{table}
\begin{center}
\begin{tabular}{lclcl}
Matter & Equation & Homology & $U(1)_X$ \\
\hline
$16_{t_{1,2}}$ &  $a_1$ & $\eta - 2 c_1 - \tilde{\chi}$  & $M-P$ \\
$16_{t_{3}}$ &  $a_4$ & $-c_1 + \chi_5$ & $P_5$ \\
$16_{t_{4}}$ &  $a_6$ & $-c_1 + \chi_7$ & $P_7$ \\
$10_{(t_1 + t_3 )}$ &  $a_1 - \gamma a_4 a_6$ & $\eta - 2 c_1 - \tilde{\chi}$ & $M-P$ \\
$10_{(t_1 + t_2 )}$ &  $a_5 a_6 + a_4 a_7$ & $-c_1 + \tilde{\chi}$ & $P$ \\
$10_{(t_1 + t_4 )}$ &  $a_1 - \gamma a_4 a_6$ & $\eta - 2 c_1 - \tilde{\chi}$ & $M-P$ \\
$10_{(t_3 + t_4 )}$ &  $a_5 a_6 + a_4 a_7$ & $-c_1 + \tilde{\chi}$ & $P$
\end{tabular}
\end{center}
\caption{\small{16 and 10 matter curves and their equations and homology classes}}
\label{SU4table}
\end{table}

\subsubsection{Homology of the $SO(10)$ singlets}

We have already pointed out that singlet fields can play a decisive role in building
the low  energy effective model. As previously, we need to determine their
properties, their homology classes, the flatness constraints etc.

Using well known theorems (see Appendix A)
the singlets are given by the equation
\begin{equation}
P_0 = -4 b_2^3 b_3^2 - 27 b_0 b_3^4 +16 b_2^4 b_4 + 144b_0 b_2 b_3^2 b_4 - 128b_0 b_2^2 b_4^2 + 256 b_0^2 b_4^3 = 0\label{su4s}
\end{equation}

\noindent When the $b_i$ are expressed in terms of the $a_i$, the results are

\begin{align}
b_0 &= \gamma (a_5 a_7 )^2 \\
b_1 &= 0 \\
b_2 &= a_5 a_7 (a_1 + \gamma a_4 a_6 )- \gamma (a_5 a_6 + a_4 a_7 )^2 \\
b_3 &= (a_1 - \gamma a_4 a_6 )(a_5 a_6 + a_4 a_7 ) \\
b_4 &= a_1 a_4 a_6
\end{align}

\noindent Factorisation of $P_0$ leads to

\begin{align}
P_0 &= [(a_5 a_6 +a_4 a_7)^2 \gamma -4 a_1 a_5 a_7] \times [a_1 a_7 - a_6 (a_2 -\gamma a_5 a_6 )]^2 \notag \\
& \times [a_1 a_5 - a_4 (a_2 - \gamma a_4 a_7 )]^2 \times [a_5 a_6 - a_4 a_7 ]^2
\end{align}

\noindent As we know the homologies of the $a_i$, we have the homologies of the singlet curves, which are summarised in Table \ref{SU4singlets}.

\begin{table}[tbp] \centering%
\begin{tabular}{|l|c|c|c|c|}
\hline
Matter&Charge& Equation& Homology&$U(1)_X$\\
\hline
$\theta_{12}$, $\theta_{21}$& $\pm (t_1 -t_2 )$& $(a_5 a_6 +a_4 a_7)^2 \gamma -4 a_1 a_5 a_7$&$\eta - 2c_1$&$M-C$\\ \hline
$\theta_{13}$, $\theta_{31}$& $\pm (t_{1,2} -t_3 )$& $a_1 a_5 - a_4 (a_2 - \gamma a_4 a_7 )$&$\eta-2c_1 -\chi_7$ &$M-P_7$\\ \hline
$\theta_{14}$, $\theta_{41}$& $\pm (t_{1,2} -t_4 )$& $a_1 a_7 - a_6 (a_2 -\gamma a_5 a_6 )$&$\eta-2c_1 -\chi_5$ &$M-P_5$\\ \hline
$\theta_{34}$, $\theta_{43}$& $\pm (t_3 -t_4)$&$a_5 a_6 - a_4 a_7 $ &$-c_1 + \chi$ & $M+P$ \\ \hline
\end{tabular}%
\caption{\small{SU(4) cover singlets and homologies}}
\label{SU4singlets}
\end{table}

\subsection{$SU(5)_{\perp}$ Spectral Cover}

Our final  investigation in the present work concerns the  $SU(5)_{GUT}$. Considering
again the maximal symmetry $E_8$, the spectral cover encoding the relevant information
(bundle structure etc) is associated to the commutant of the GUT group, which is
$SU(5)_{\perp}$.  Hence, in this case the breaking pattern is
\begin{equation}
 E_8 \rightarrow SU(5) \times SU(5)_{\perp}
\end{equation}
This case has been extensively studied and the homology of the gauge non-singlets determined.  Here we extend the discussion to include the singlets.
The associated adjoint representation decomposition is
\begin{equation}
 248  \rightarrow (24,1)+(1,24) +(10,5)+(\overline{5},10)+(\overline{5}, \overline{10})+(5,\overline{10})
 \end{equation}
Although this case has been analysed by many authors in the recent
F-theory model building literature, a detailed examination of the  breaking
mechanism  of the higher intermediate  symmetries and possible implications
is still lacking. In the following we attempt to implement the constraints
obtained from the previous symmetry breaking stages into the $SU(5)_{GUT}$
model.

To start with, we recall that the global model is assumed in the context
of  elliptically fibered Calabi-Yau compact complex fourfold over
a three-fold base. Using the Tate's algorithm\cite{Tate1975,Bershadsky:1996nh},
the ${\bf SU(5)}$ singularity can be described by the following form of
 Weierstrass' equation ~\cite{Donagi:2008ca}
 \ba
 y^2=x^3+b_0 { z^5}+b_2 x { z^3}+b_3yz^2+b_4x^2 { z}+b_5xy\nn
 \ea
We determine the corresponding spectral cover by defining homogeneous coordinates
\[z\ra U,\; x\ra V^2,\; y\ra V^3\]
so that the spectral cover equation becomes
\ba
 0&=&b_0 { U^5}+b_2 V^2 { U^3}+b_3V^3U^2+b_4V^4 {U}+b_5V^5\nn
 \ea
We can see this equation as a fifth degree polynomial in terms of  the affine parameter
$s=U/V$:
\[P_5=\sum_{k=0}^5 b_ks^{5-k}\;=\;b_5+b_4s+b_3s^2+b_2s^3+b_1s^4+b_0s^5\]
where we have divided by $V^5$, so that each term in the
last equation becomes section of $c_1-t$.
The roots of the spectral cover equation.
\[0=b_5+b_4s+b_3s^2+b_2s^3+b_0s^5\propto \prod_{i=1}^5(s+t_i)\]
are identified as the $SU(5)$ weights  $t_i$.

In the above the coefficient $b_1$ is taken to be zero since it corresponds to the sum
of the roots which for $SU(n)$ is always zero, $\sum t_i=0$.
Also, it can be seen that the coefficient $b_5$ is equal to the product of
the roots, i.e. $b_5=t_1t_2t_3t_4t_5$ and the  $\Sigma_{10}$ curves where
the corresponding matter multiplets are  localized are determined by
the five zeros
\ba
\Sigma_{10_i},\;\;\; b_5=\prod_{i=1}^5t_i=0\ra t_i=0,\;\;\; i=1,2,3,4,5\label{tens}
\ea
Following~\cite{Dudas:2010zb}, we impose the  ${\cal Z}_2$  monodromy corresponding to the
following splitting of the spectral cover equation
\ba
0&=&(a_1+a_2s+a_3s^2)(a_4+a_5s)(a_6+a_7s)(a_8+a_9s)\label{4split}
\ea
with $s=U/V$ and $a_i$ undetermined coefficients, constituting sections of line bundles
each of them being of  specific  Chern class to be determined.
 The first bracket contains the polynomial factor which corresponds to the  ${\cal Z}_2$
monodromy, while the remaining monomials leave three $U(1)$'s intact.
Expanding, we may determine the homology class for each of the coefficients $a_i$ by comparison with
the $b_k$'s. Thus,
\ba
b_0&=&a_3 a_5 a_7 a_9\nn
\\
b_1&=& a_3 a_5 a_7 a_8+a_3 a_4 a_9 a_7+a_2 a_5 a_7 a_9 +a_3 a_5 a_6 a_9\nn\\
b_2&=&a_3 a_5 a_6 a_8+a_2 a_5 a_8 a_7+a_2 a_5 a_9 a_6+a_1 a_5 a_9 a_7+a_3 a_4 a_7 a_8+a_3 a_4 a_6
   a_9+a_2 a_4 a_7 a_9\nn\\
   b_3&=&a_3 a_4 a_8 a_6+a_2 a_5 a_8 a_6+a_2 a_4 a_8 a_7+a_1 a_7 a_8 a_5+a_2 a_4 a_6 a_9+a_1 a_5 a_6  a_9+a_1 a_4 a_7 a_9\nn\\
   b_4&=&a_2 a_4 a_8 a_6+a_1 a_5 a_8 a_6+a_1 a_4 a_8 a_7+a_1 a_4 a_6 a_9\nn\\
   b_5&=&a_1a_4a_6a_8\nn
   \ea
We first solve the constraint $b_1=0$.  We make the Ansatz:
\[ a_2 =-c ( a_5 a_7 a_8+a_4 a_9 a_7+ a_5 a_6 a_9),\; a_3=c  a_5 a_7 a_9\]
Substituting into $b_n$'s we get
\ba
b_0&=&c\, a_5^2 a_7^2 a_9^2 \notag \\
b_2&=&a_1 a_5 a_7 a_9-\left(a_5^2 a_7^2 a_8^2+a_5 a_7 \left(a_5 a_6+a_4 a_7\right) a_9 a_8+\left(a_5^2 a_6^2+a_4 a_5 a_7 a_6+a_4^2 a_7^2\right) a_9^2\right) c \notag
\\
b_3&=&a_1 \left(a_5 a_7
a_8+a_5 a_6 a_9+a_4 a_7 a_9\right)-\left(a_5 a_6+a_4 a_7\right) \left(a_5 a_8+a_4 a_9\right) \left(a_7 a_8+a_6 a_9\right) c \notag \\
b_4&=&a_1 \left(a_5 a_6 a_8+a_4 a_7 a_8+a_4 a_6 a_9\right)-a_4 a_6 a_8 \left(a_5 a_7 a_8+a_5 a_6 a_9+a_4 a_7 a_9\right) c \notag \\
b_5&=&a_1 a_4 a_6 a_8 \notag
\ea

\begin{table}[tbp] \centering%
\begin{tabular}{|l|c|c|c|c|c|}
\hline
Matter&Charge& Equation& Homology&$N_Y$&$M_{U(1)}$\\
\hline
$5_{H_u}$& $-2 t_{1}$& $a_8 a_5 a_7 + a_6 a_5 a_9 + a_4 a_7 a_9$&$-c_1 +\tilde{\chi}$ &$\tilde{N}$&$M_{H_u}$\\ \hline
$5_1$& $-t_1 - t_{3}$& $a_1 -ca_4 a_8 a_7 -ca_4 a_6 a_9$&$\eta-2c_1 -\tilde{\chi}$ &$-\tilde{N}$&$M_{5_1}$\\ \hline
$5_2$& $-t_1 - t_{4}$& $a_1 -ca_6 a_8 a_5 -ca_4 a_6 a_9$&$\eta-2c_1 -\tilde{\chi}$ &$-\tilde{N}$&$M_{5_2}$\\ \hline
$5_3$& $-t_1 - t_{5}$& $a_1 -ca_6 a_8 a_5 -ca_4 a_8 a_7$&$\eta-2c_1 -\tilde{\chi}$ &$-\tilde{N}$&$M_{5_3}$\\ \hline
$5_4$& $-t_{3}-t_4$& $a_6 a_5 +a_4 a_7$&$-c_1 + \chi_5 +\chi_7$ &$N_5 +N_7$&$M_{5_4}$\\ \hline
$5_5$& $-t_{3}-t_5$& $a_8 a_5 +a_4 a_9$&$-c_1 + \chi_5 +\chi_9$ &$N_5 +N_9$&$M_{5_5}$\\ \hline
$5_6$& $-t_{4}-t_5$& $a_8 a_7 +a_6 a_9$&$-c_1 + \chi_7 +\chi_9$ &$N_7 +N_9$&$M_{5_6}$\\ \hline
$10_M$& $t_1$& $a_1$&$\eta-2c_1 -\tilde{\chi}$ &$-\tilde{N}$&$-(M_{5_1}+M_{5_2}$\\
  &   &   &  &  &$+M_{5_3})$\\ \hline
$10_2$& $t_3$& $a_4$&$-c_1 + \chi_5$ &$N_5$&$M_{10_2}$\\ \hline
$10_3$& $t_4$& $a_6$&$-c_1 + \chi_7$ &$N_7$&$M_{10_3}$\\ \hline
$10_4$& $t_5$& $a_8$&$-c_1 + \chi_9$ &$N_9$
&$M_{10_4}$\\ \hline
$\theta_{12}$&$t_1 -t_2$&$2 a_5 a_7 a_9 \left(a_4 \left(a_7 a_8+a_6 a_9\right) c-2 a_1\right)$&$\eta-2c_1$&$0$&$M_{12}$\\
  &   &$+a_4^2 a_7^2 a_9^2 c+a_5^2 \left(a_7 a_8+a_6 a_9\right){}^2 c$&  &  &\\ \hline
$\theta_{13}$&$t_1 -t_3$&$a_4 \left(2 a_4 a_7 a_9+a_5 \left(a_7 a_8+a_6 a_9\right)\right) c$&$\eta-2c_1-\tilde{\chi}+\chi_5$&0&$M_{13}$\\
  &   &$+a_1 a_5$&  &  &\\ \hline
$\theta_{14}$&$t_1 -t_4$&$ a_6 \left(a_4 a_7 a_9+a_5 \left(a_7 a_8+2 a_6 a_9\right)\right) c$&$\eta-2c_1-\tilde{\chi}+\chi_7$&0&$M_{14}$\\
  &   &$+a_1 a_7$&  &  &\\ \hline
$\theta_{15}$&$t_1 -t_5$&$2 a_5 a_7 a_8^2 c+a_9 a_8  \left(a_5 a_6+a_4 a_7\right)  c $&$\eta-2c_1-\tilde{\chi}+\chi_9$&0&$M_{15}$\\
  &   &$+a_1 a_9$&  &  &\\ \hline
$\theta_{34}$&$t_3 -t_4$&$a_5 a_6-a_4 a_7 $&$-c_1+\chi_5+\chi_7$&0&$M_{34}$\\ \hline
$\theta_{35}$&$t_3 -t_5$&$a_5 a_8-a_4 a_9$&$-c_1+\chi_5+\chi_9$&0&$M_{35}$\\ \hline
$\theta_{45}$&$t_4 -t_5$&$a_7 a_8-a_6 a_9 $&$-c_1+\chi_7+\chi_9$&0&$M_{45}$\\ \hline
$\theta_0$&$-$     &$ca_5^2 a_7^2a_9^2$&$\eta$&$0$&$M_0$\\ \hline
\end{tabular}%
\caption{\small{Table showing curves and flux restrictions with  ${\cal Z}_2$  monodromy. $\tilde{N}=N_5 +  N_7 + N_9$.}}
\label{DPtable}
\end{table}

\begin{table}[tbp] \centering%
\begin{tabular}{|l|c|c|c|}
\hline
Singlet&$Q_\chi$& $Q_{\psi}$& Representations\\
\hline
$\theta_{12}$& 0 & 0 & SO(10) singlet in 78 \\ \hline
$\theta_{13}$&0 & 0 & $45 \subset 78$\\ \hline
$\theta_{14}$& 0 & 4 & SO(10) singlet in $27_{t_{1,2}}$\\ \hline
$\theta_{15}$& -5 & 1 & $16_{t_{1,2}} \subset 27_{t_{1,2}}$ \\ \hline
$\theta_{34}$& 0 & 4 & SO(10) singlet in $27_{t_{3}}$\\ \hline
$\theta_{35}$& -5 & -1 & $16_{t_{3}} \subset 27_{t_{3}}$ \\ \hline
$\theta_{45}$& -5 & -3 & $16_{t_{4}} \subset 78$\\ \hline
\end{tabular}
\caption{\small{Table showing the $E_6$ charges and origin of some of the singlets in Table~\ref{DPtable}.}}
\label{DPtable0}
\end{table}

Next, we observe that we have  to determine the homology classes of nine unknowns $a_1,\dots a_9$
   in terms of the $b_k$-classes, which we demand to be $\eta - k c_1$. Three classes are left  unspecified which we
   choose them to be $[a_l]=\chi_l,l=5,7,9$. The rest are computed easily, and the results are $[a_1 ]=\eta-2c_1-\chi$, $[a_2 ]=\eta-c_1-\chi$, $[a_3 ]=\eta-\chi$, $[a_4 ]=-c_1+\chi_5$, $[a_5 ]=\chi_5$, $[a_6 ]=-c_1+\chi_7$, $[a_7 ]=\chi_7$, $[a_8 ]=-c_1+\chi_9$, $[a_9 ]=\chi_9$.

The $\Sigma_{10}$ curves are found setting $s=0$ in the polynomial
\ba b_5\equiv\Pi_5(0)=a_1a_4a_5a_6=0&\ra& a_1=0,a_4=0,a_5=0,a_6=0\ea
Thus, after the monodromy action, we obtain four curves (one less)
to arrange the appropriate pieces of the three (3) families.

The $\Sigma_{5}$ curves are treated similarly in~\cite{Dudas:2010zb} so we do not present the details here.

\subsection{Singlets in the $SU(5)_{\perp}$ Spectral Cover}

For the singlets, we need to determine the polynomial   $\prod_{i\ne j} (t_i-t_j)$ in terms of $b_n$'s.   In analogy with the previous cases, while using the results of Appendix A
we find
\ba
P_0&=&3125 b_5^4 b_0^5+256 b_4^5 b_0^4-3750 b_2 b_3 b_5^3 b_0^4+2000 b_2 b_4^2 b_5^2 b_0^4+2250 b_3^2 b_4 b_5^2 b_0^4\nn\\
                &&-1600 b_3 b_4^3 b_5 b_0^4-128 b_2^2 b_4^4 b_0^3+144 b_2 b_3^2 b_4^3 b_0^3-27 b_3^4 b_4^2 b_0^3+825 b_2^2 b_3^2 b_5^2 b_0^3\nn\\
                &&-900 b_2^3 b_4 b_5^2 b_0^3+108 b_3^5 b_5   b_0^3+560 b_2^2 b_3 b_4^2 b_5 b_0^3-630 b_2 b_3^3 b_4 b_5 b_0^3\nn\\
                &&+16 b_2^4 b_4^3 b_0^2-4 b_2^3 b_3^2 b_4^2 b_0^2+108 b_2^5 b_5^2 b_0^2+16 b_2^3 b_3^3 b_5 b_0^2-72 b_2^4 b_3 b_4 b_5 b_0^2\nn
\ea
Factorization (via Mathematica) leads to the results which are summarised in the complete SU(5) table (Table  \ref{DPtable}). Since the factorised form is very lengthy we do not exhibit it here. {Note that due to the large number of parameters $\eta,\;c_{1},\;\tilde{\chi},\;\tilde{\chi}_{5,7,9}$ there are no constraints between the singlet $M_{U(1)}$s for the case that only the hyperchage is unbroken by flux effects.}

The way in which the singlets fit into the $E_6$ and SO(10) pictures can be found by working out the $U(1)_\chi$ and $U(1)_\psi$ charges using the generators in Eq. (\ref{charges}) and
 matching the charges to the singlets in the decomposition in Eq. (\ref{E601}).
Putting this information together with the homology classes, leads to the results in Tables~\ref{DPtable} and
\ref{DPtable0}.

\section{Singlet VEVs and D- and F-flatness conditions}

The homology constraints just discussed can be used to construct models capable of accommodating the Standard Model - an example of this is given in Section \ref{sec:model}. To obtain a viable model it is usually necessary to remove additional Standard Model ``vectorlike'' states by generating mass for them through their coupling to $E_{6}$ singlets which acquire VEVs. Any such VEVs  should be consistent with F and D flatness conditions and we turn now to a discussion of this.  Since, in this paper, we have assumed all GUT breaking is driven by flux no GUT non-singlet fields acquire VEVs until the electroweak scale and so these VEVs can be ignored when determining high scale VEVs.

In general the superpotential for the massless singlet fields is given by
\begin{equation}
W = \mu_{ijk} \theta_{ij} \theta_{jk} \theta_{ki}
\end{equation}

\noindent The F-flatness conditions are given by

\begin{equation}
\frac{\partial W}{\partial \theta_{ij}} = \mu_{ijk} \theta_{jk} \theta_{ki} = 0
\end{equation}

\noindent The D-flatness condition for $U_{A}(1)$ is

\begin{equation}
\sum_{j} Q^A_{ij} (\left|\left\langle \theta_{ij} \right\rangle \right|^2 -\left|\left\langle \theta_{ji} \right\rangle \right|^2 ) = - \frac{Tr Q^A}{192 \pi^2} g_s^2 M_S^2
\label{Dflat}
\end{equation}
where the right-hand side (rhs) is the anomalous contribution, $Q^{A}_j$ are the singlet charges and the trace $TrQ^A$ is over all singlet and non-singlet states. The D-flatness conditions must be checked for each of the $U_{A}(1)$s.

\subsection{$E_{6}$ case}
In this case after the monodromy action there is only a single $U(1)$ and, in the $t_{i}'$ basis the charge is given by diag$[1,1,-2]$.
As both the 27s and the $\theta_{ij}$ are charged under the U(1), we must know the number of each after the monodromy action and the flux breaking mechanism in order to compute the trace.  The contribution of the $27_{t_i}$ to $Tr Q^A$ is

\begin{equation}
27(q_1 n_1 + q_3 n_3)=27(n_1 -2n_3 )q_1
\end{equation}

\noindent and the contribution of the $\theta_{ij}$ is

\begin{equation}
1 \times [(q_1 -q_2 )n_{12}+(q_1 -q_3 )n_{13}]=3 n_{13} q_1
\end{equation}

\noindent The multiplicities are given in terms of the flux restrictions as the flux dotted with the homology class, and so we have

\begin{align}
n_1 +n_3 &= \mathcal{F} \cdot (\eta - 3 c_1) \\
n_{12} = n_{13} &= \mathcal{F} \cdot (\eta - 2 c_1)
\end{align}

Assuming that only the pair $\theta_{13}$, $\theta_{31}$ get VEVs, the flatness condition is

\begin{equation}
q_3 (\left|\left\langle \theta_{13} \right\rangle \right|^2 -\left|\left\langle \theta_{31} \right\rangle \right|^2 ) + \frac{9(n_1 -2 n_3 )+n_{13}}{64 \pi^2} q_1 g_s^2 M_S^2 = 0
\end{equation}

\noindent and as we have $q_3 = -2q_1$

\begin{equation}
\left|\left\langle \theta_{13} \right\rangle \right|^2 -\left|\left\langle \theta_{31} \right\rangle \right|^2 = \frac{9(n_1 -2 n_3 )+n_{13}}{128 \pi^2} g_s^2 M_S^2
\end{equation}

In order to relate the multiplicities to each other, we define for convenience $\omega = \mathcal{F} _{U(1)} \cdot \eta$, $p = \mathcal{F} _{U(1)} \cdot c_1$ and $x = \mathcal{F} _{U(1)} \cdot \chi$.  As such, in this notation, we have

\begin{align}
n_1 &= \omega -2p -x \\
&= n_{13} -x \\
n_3 &= x - p
\end{align}

\noindent As chirality requires $n_1 > 0$ and $n_3 < 0$, the term $9(n_1 -2 n_3 )$ is always positive.  If we take the case $n_1 = 4$ and $n_3 = -1$ (i.e. the minimal case of three
$27$'s accommodating the three families and a pair $27_{H}+\overline{27}_{\bar H}$), we have $n_{13}=3+p$, and

\begin{equation}
\left|\left\langle \theta_{13} \right\rangle \right|^2 -\left|\left\langle \theta_{31} \right\rangle \right|^2 = \frac{54+n_{13}}{128 \pi^2} g_s^2 M_S^2
\end{equation}

\noindent This condition is consistent with $\left\langle \theta_{13} \right\rangle \neq 0$ and $\left\langle \theta_{31} \right\rangle = 0$ for any $n_{13} > 0$, but not with the case $\left\langle \theta_{13} \right\rangle =0$, $\left\langle \theta_{31} \right\rangle \neq 0$ as this would require $n_{13} < -54$.

\subsection{SO(10) case}

Analogous to the $E_6$ case, the D-flatness condition for the anomalous U(1)s is given by Eq. (\ref{Dflat}). In this case there are two $U_{A}(1)$s with charges that can be taken as $Q^{1}={\rm diag}[1,1,1,-3]$ and $Q^{2}={\rm diag}[1,1,-2,0]$.
For example, for the case of $Q_{1}$, using Table \ref{SU4table}, the trace  is given by

\begin{align}
Tr Q^1_j &= 16 (n^{16}_1 + n^{16}_3  -3n^{16}_4) + 10(2n^{10}_{13}+2n^{10}_{12}-2n^{10}_{14}-2n^{10}_{34})+
4n^{1}_{14}+4n^{1}_{34}
\end{align}

\subsection{SU(5) case}
In this case there are three $U_{A}(1)$s with charges given in Eq. (\ref{charges}). In the next Section we discuss F- and D-flatness in detail for a realistic model.

\section{Model Building: a realistic model based on $E_{6}$}\label{sec:model}
There are several important ingredients to building a phenomenologically realistic low energy theory. The first is the need to control the baryon- and lepton-number violating terms in the Lagrangian that generate rapid nucleon decay. In addition to the dimension 3 and 4 terms (forbidden by R-parity in the MSSM) it is necessary to forbid the dimension 5 nucleon decay terms too. Although the latter are suppressed by an inverse mass factor, this mass must be some $10^{7}$ times the Planck mass, unacceptably large in string theory.

A second necessary ingredient is the control of the ``$\mu$ term''. The Higgs doublet supermultiplet mass term in the superpotential, $\mu H_{u}H_{d}$, is allowed by the Standard Model Gauge symmetry but, for a viable theory,  its coefficient, $\mu$, must be of order the SUSY breaking scale in the visible sector. To control this requires additional symmetry. At the same time the Higgs colour triplets that are expected as partners of the Higgs doublets in a unified theory must be very heavy - the ``doublet-triplet splitting'' problem.

Finally the quark and lepton masses and mixings must be consistent. In particular it is necessary to explain why the quark masses and mixing angles have a hierarchical structure while the leptons must have large mixing angles and a relatively small mass hierarchy to explain the observed neutrino oscillation phenomena.

There has been a significant effort to build  F-theory based models that use its $U(1)$ symmetries to obtain these ingredients but, to date, no fully satisfactory model has been obtained; indeed it has been speculated that it is not possible. Here, using the results  obtained above, we construct an explicit example to demonstrate how the $U(1)$ symmetries alone are sufficient to build a viable theory.
\subsection{The $E_{6}$ inspired model}
The first, most important, step in model building is to find a matter and Higgs multiplet assignment that can eliminate rapid nucleon decay. In this we find that starting from an underlying unified group is very helpful and we consider the case of $E_{6}$. After imposing a ${\cal Z}_2$ monodromy there are just two multiplets, $27_{t_{1,3}'}$. The $SU(5)\times SU(5)_{\perp}$ properties of these multiplets are given in Table \ref{t1}. The only $E_{6}$ allowed trilinear term in the superpotential is $27_{t_{1}'}27_{t_{1}'}27_{t_{3}'}$.  As a result, if we assign the quark and lepton supermultiplets to $27_{t_{1}}$ and the Higgs supermultiplets to $27_{t_{3}}$, there will be no dimension 3 or dimension 4 baryon- or lepton-number violating terms.

Anomaly cancellation leads to constraints between the number of $SU(5)$ 10 and 5 dimensional representations~\cite{Dudas:2010zb,Marsano:2010sq}.  These conditions are automatically satisfied for multiplets descending from complete $E_{6}$ multiplets. In particular for the $E_{6}$ 27 dimensional representations we have, in the notation of~\cite{Dudas:2010zb}
\bea
M_{10_M}=M_{5_1}=-M_{5_2}=-M_{5_3},  \\
M_{10_2}=-M_{5_4}=-M_{5_5}=M_{5_{H_u}}.
\eea
Furthermore, in the absence of matter in the 78 dimensional representation we have
\be
M_{10_3}=M_{10_4}=M_{5_6}=N_8=N_9=0,
\ee
which implies:
\be
N_7=\tilde{N}.
\ee
The resulting states arising from complete 27s are shown in Table~\ref{no1} where we have allowed also for the breaking of $SU(5)$ through hypercharge flux. The SM particle content is also shown in Table~\ref{no1} in the usual notation where
a generation of quarks and leptons is $Q,u^c,d^c,L,e^c$.
The Higgs doublets $H_u,H_d$ are accompanied by exotic colour triplets
and anti-triplets $D,\overline{D}$. The 27s also contain the CP conjugates
of the right-handed neutrinos $\nu^c$ and extra singlets $S$.

\begin{table}[htdp]
\begin{center}
\small
\begin{tabular}{|c|c|c|c|c|c|c|}
\hline
$E_6$ & $SO(10)$ & $SU(5)$  & Weight vector & $N_Y$ & $M_{U(1)}$ & SM particle content\\
\hline
$27_{t_1'}$ & $16$ & $\overline{5}_3$ & $t_1+t_5$ & $\tilde{N}$ & $-M_{5_3}$ &
$-M_{5_3}d^c+(-M_{5_3}+\tilde{N})L$\\
\hline
$27_{t_1'}$ & $16$ & $10_M$ & $t_1$ & $-\tilde{N}$ & $-M_{5_3}$ &
$-M_{5_3}Q+(-M_{5_3}+\tilde{N})u^c+(-M_{5_3}-\tilde{N})e^c$\\
\hline
$27_{t_1'}$ & $16$ & $\theta_{15}$ & $t_1-t_5$ & $0$ & $-M_{5_3}$ &
$-M_{5_3}\nu^c$\\
\hline
$27_{t_1'}$ & $10$ & $5_1$ & $-t_1-t_3$ & $-\tilde{N}$ & $-M_{5_3}$ &
$-M_{5_3}D+(-M_{5_3}-\tilde{N})H_u$\\
\hline
$27_{t_1'}$ & $10$ & $\overline{5}_2$ & $t_1+t_4$ & $\tilde{N}$ & $-M_{5_3}$ &
$-M_{5_3}\overline{D}+(-M_{5_3}+\tilde{N})H_d$\\
\hline
$27_{t_1'}$ & $1$ & $\theta_{14}$ & $t_1-t_4$ & $0$ & $-M_{5_3}$ &
$-M_{5_3}S$\\
\hline
$27_{t_3'}$ & $16$ & $\overline{5}_5$ & $t_3+t_5$ & $-\tilde{N}$ & $M_{5_{H_u}}$ &
$M_{5_{H_u}}d^c+(M_{5_{H_u}}-\tilde{N})L$\\
\hline
$27_{t_3'}$ & $16$ & $10_2$ & $t_3$ & $\tilde{N}$ & $M_{5_{H_u}}$ &
$M_{5_{H_u}}Q+(M_{5_{H_u}}-\tilde{N})u^c+(M_{5_{H_u}}+\tilde{N})e^c$\\
\hline
$27_{t_3'}$ & $16$ & $\theta_{35}$ & $t_3-t_5$ & $0$ & $M_{5_{H_u}}$ &
$M_{5_{H_u}}\nu^c$\\
\hline
$27_{t_3'}$ & $10$ & $5_{H_u}$ & $-2t_1$ & $\tilde{N}$ & $M_{5_{H_u}}$ &
$M_{5_{H_u}}D+(M_{5_{H_u}}+\tilde{N})H_u$\\
\hline
$27_{t_3'}$ & $10$ & $\overline{5}_4$ & $t_3+t_4$ & $-\tilde{N}$ & $M_{5_{H_u}}$ &
$M_{5_{H_u}}\overline{D}+(M_{5_{H_u}}-\tilde{N})H_d$\\
\hline
$27_{t_3'}$ & $1$ & $\theta_{34}$ & $t_3-t_4$ & $0$ & $M_{5_{H_u}}$ &
$M_{5_{H_u}}S$\\
\hline
\end{tabular}
\end{center}
\caption{\small Complete $27$s of $E_6$ and their $SO(10)$ and $SU(5)$ decompositions.
 The indices of the $SU(5)$ non-trivial states $10,5$ refer to the labeling of the
corresponding  matter curve (we use the notation of~\cite{Dudas:2010zb}). We impose the
extra conditions on the integers $N_Y$ and $M_{U(1)}$ from the requirement of having complete
27s of $E_6$ and no 78 matter. The $SU(5)$ matter states decompose into SM states as
$\overline{5}\rightarrow d^c,L$ and $10\rightarrow Q,u^c,e^c$ with right-handed neutrinos
$1\rightarrow \nu^c$, while $SU(5)$ Higgs states decompose as $5\rightarrow D,H_u$ and
$\overline{5}\rightarrow \overline{D},H_d$, where $D, \overline{D}$ are exotic colour triplets and antitriplets.
We identify RH neutrinos as $\nu^c=\theta_{15,35}$ and extra singlets from the 27 as $S=\theta_{14,34}$.}
\label{no1}
\end{table}%

The only undetermined parameters in Table~\ref{no1} are the three integers $M_{5_3}$, $M_{5_{H_u}}$ and
$\tilde{N}$. To maintain the $E_{6}$ based suppression of the baryon- and lepton-number violating terms we require that the Higgs should come from $27_{t_{3}'}$ and the matter from $27_{t_{1}'}$ and that any states transforming as $H_{u,d}$ in $27_{t_{1}'}$ be heavy.

We first choose $M_{5_3}=-3$ to get three families of quarks and leptons in $27_{t_{1}'}$. To get a single pair of Higgs doublets in $27_{t_{3}'}$ without colour triplet partners we next choose $M_{5_{H_u}}=0$ and $\tilde{N}=1.$
According to Table~\ref{no1} this gives the following SM spectrum, grouped according to $SO(10)$ origin:
\bea
&&\left[\overline{5}_3\rightarrow 3d^c+4L, \
10_M\rightarrow 3Q+4u^c+2e^c, \
\theta_{15}\rightarrow 3\nu^c\right]_{16}, \nonumber \\
&&\left[5_1\rightarrow 3D+2H_u, \
\overline{5}_2\rightarrow 3\overline{D}+4H_d\right]_{10}, \nonumber \\
&&\left[\theta_{14}\rightarrow 3S\right]_{1}, \nonumber \\
&&\left[\overline{5}_5\rightarrow \overline{L}, \
10_2\rightarrow \overline{u^c}+e^c\right]_{16}, \nonumber \\
&&\left[5_{H_u}\rightarrow H_u, \
\overline{5}_4\rightarrow \overline{H_d}\right]_{10}. \nonumber \\
\eea
Note that the matter content is just that contained in 3 complete 27s of $E_6$:
$3[Q,u^c,d^c,L,e^c,\nu^c]_{16}$, $3[H_u,D,H_d,\overline{D}]_{10}$, $3[S]_1$
plus some extra vector pairs $L+\overline{L}, e^c+\overline{e^c}, u^c+\overline{u^c},H_d+\overline{H_d}$ that may be expected to get a large mass if some of the singlet states acquire large VEVs.

It may be seen that the $U(1)$ flux breaking has resulted in one of the lepton supermultiplets, $e^{c}$, being assigned to $27_{t_{3}'}$ in conflict with our original strategy of assigning all matter states to $27_{t_{1}'}$. However this does not lead to the dimension 4 R-parity violating superpotential term $LLe^{c}$  because one of the $e^{c}$ comes from the $16$ of $SO(10)$ and there is no $16^{3}$ coupling allowed by $SO(10)$. In this case it is a combination of the original R-parity and the underlying GUT symmetry that eliminates dangerous baryon- and lepton-number violating terms. In fact the combination is more effective than R-parity alone for it also forbids the dangerous dimension 5 terms.

More troublesome is the fact that $H_{d}$ now comes from $27_{t_{1}'}$ so that down quark masses are forbidden in tree level. However there is an allowed coupling of  $H_{d}Le^{c}$ for the $e^{c}$ belonging to $27_{t_{3}'}$. This discrepancy between down quark and charged lepton masses looks unacceptable even if the remaining masses are generated in higher order through coupling to singlet fields that acquire large VEVs. To avoid this we look at a slightly modified structure choosing
\bea
M_{10_M}&=&-M_{5_3}=4, \nonumber \\
M_{5_1}&=&-M_{5_2}=3\nonumber\\
M_{10_2}&=&-M_{5_{5}}=-1,\nonumber\\
M_{5_4}&=&M_{H_{u}}=0,\nonumber\\
M_{\theta_{15}}&=&2,\nonumber\\
\tilde{N}&=&1
\eea
This leads to the spectrum given in Table \ref{2}
where now both the down quarks and leptons originate in $27_{t_{1}'}$, avoiding the troublesome difference in their mass matrices just discussed.

\begin{table}[htdp]
\begin{center}
\small
\begin{tabular}{|c|c|c|c|c|c|c|c|}
\hline
$E_6$ & $SO(10)$ & $SU(5)$  & Weight vector & $N_Y$ & $M_{U(1)}$ & SM particle content& Low energy spectrum\\
\hline
$27_{t_1'}$ & $16$ & $\overline{5}_3$ & $t_1+t_5$ & $1$ & $4$ &
$4d^c+5L$&$3d^{c}+3L$\\
\hline
$27_{t_1'}$ & $16$ & $10_M$ & $t_1$ & $-1$ & $4$ &
$4Q+5u^c+3e^c$&$3Q+3u^{c}+3e^{c}$\\
\hline
$27_{t_1'}$ & $16$ & $\theta_{15}$ & $t_1-t_5$ & $0$ & $3$ &
$3\nu^c$&-\\
\hline
$27_{t_1'}$ & $10$ & $5_1$ & $-t_1-t_3$ & $-1$ & $3$ &
$3D+2H_u$&-\\
\hline
$27_{t_1'}$ & $10$ & $\overline{5}_2$ & $t_1+t_4$ & $1$ & $3$ &
$3\overline{D}+4H_d$&$H_{d}$\\
\hline
$27_{t_3'}$ & $16$ & $\overline{5}_5$ & $t_3+t_5$ & $-1$ & $-1$ &
$\overline{d^c}+2\overline{L}$&-\\
\hline
$27_{t_3'}$ & $16$ & $10_2$ & $t_3$ & $1$ & $-1$ &
$\overline{Q}+2\bar{u^c}$&-\\
\hline
$27_{t_3'}$ & $16$ & $\theta_{35}$ & $t_3-t_5$ & $0$ & $0$ &
$-$&-\\
\hline
$27_{t_3'}$ & $10$ & $5_{H_u}$ & $-2t_1$ & $1$ & $0$ &
$H_u$&$H_{u}$\\
\hline
$27_{t_3'}$ & $10$ & $\overline{5}_4$ & $t_3+t_4$ & $-1$ & $0$ &
$\overline{H_d}$&-\\
\hline
$27_{t_3'}$ & $1$ & $\theta_{34}$ & $t_3-t_4$ & $0$ & $1$ &
$\theta_{34}$&-\\
\hline
- & $1$ & $\theta_{31}$ & $t_3-t_1$ & $0$ & $4$ &
$\theta_{31}$&-\\
\hline
- & $1$ & $\theta_{53}$ & $t_5-t_3$ & $0$ & $1$ &
$\theta_{53}$&-\\
\hline
- & $1$ & $\theta_{14}$ & $t_1-t_4$ & $0$ & $3$ &
$\theta_{14}$&-\\
\hline
- & $1$ & $\theta_{45}$ & $t_4-t_5$ & $0$ & $2$ &
$\theta_{45}$&-\\
\hline
\end{tabular}
\end{center}
\caption{\small Complete $27$s of $E_6$ and their $SO(10)$ and $SU(5)$ decompositions.
We use the notation of ref~\cite{Dudas:2010zb}
for the indices of the $SU(5)$ states  and impose the
extra conditions on the integers $N_Y$ and $M_{U(1)}$ from the requirement of having complete
27s of $E_6$ and no 78 matter.
The $SU(5)$ matter states decompose into SM states as
$\overline{5}\rightarrow d^c,L$ and $10\rightarrow Q,u^c,e^c$ with right-handed neutrinos
$1\rightarrow \nu^c$, while $SU(5)$ Higgs states decompose as $5\rightarrow D,H_u$ and
$\overline{5}\rightarrow \overline{D},H_d$, where $D, \overline{D}$ are exotic colour triplets and antitriplets.
We identify RH neutrinos as $\nu^c=\theta_{15}$. The extra singlets are needed for giving mass to neutrinos and exotics and to ensure F- and D- flatness.}
\label{2}
\end{table}%

The difference in the spectrum compared to the previous case is in the vectorlike sector with additional pairs of $L+\overline{L}, Q+\overline{Q}, u^c+\overline{u^c},d^{c}+\overline{d^{c}}$ and $H_d+\overline{H_d}$ and no $e^{c}+\overline{e^{c}}$. Provided the vectorlike states are heavy the absence of the dimension 3 and 4 R-parity violating operators is now guaranteed by the underlying $U(1)$
symmetries\footnote{Note that these operators do not involve $H_{d}$ and so the fact that $H_{d}$ originates in $27_{t_{1}'}$ does not cause problems.}. As we shall see the underlying GUT symmetry still also eliminates the dimension 5 terms that would cause nucleon decay.

\subsection{Doublet-triplet splitting and vector-like masses.}

There remains the doublet-triplet problem of giving large mass to the $D$ and $\overline{D}$ fields and the problem of giving large mass to the vectorlike pairs of fields. Since the $D$ and $\overline{D}$ fields also come in vectorlike pairs these problems are related and are solved by generating mass for vectorlike fields through their coupling to SM singlet fields that acquire large VEVs. For the case the vectorlike pairs have components in both the $27_{t_{1}'}$ and $27_{t_{3}'}$ multiplets the extra vector pairs are removed by introducing $\theta_{31}$, an $E_6$ singlet, with couplings:
\beq
\theta_{31}27_{t_1'}\overline{27_{t_3'}}= \theta_{31}Q\overline{Q}+ \theta_{31}(2u^c)(2\overline{u^c})+
 \theta_{31}d^c\overline{d^c}+ \theta_{31}(2L)(2\overline{L})+
\theta_{31}H_d\overline{H_d}.
\label{31}
\eeq
If $\theta_{31}$ gets a large VEV  these vector states get large masses as required. We shall discuss how the D-terms associated with the anomalous $U_{A}(1)$s can require a VEV for this field close to the Planck scale.

To remove the remaining exotics we introduce $\theta_{34}$ which has the couplings :
\beq
\theta_{34}5_1\overline{5_2}
=\theta_{34}[3D+2H_u][3\overline{D}+3H_d]=
\theta_{34}[3(D\overline{D})]+\theta_{34}[2(H_uH_d)].
\label{34}
\eeq
If it too acquires a large VEV it generates large mass to the three copies of $D+\overline{D}$ (solving the doublet-triplet splitting problem) and two families of Higgs $H_u,H_d$, leaving just the MSSM spectrum as shown in the last column of Table \ref{2}.

\subsection{Singlet VEVS}\label{singletvevs}
In the model under consideration we assume the SUSY breaking soft masses are such that only the SM singlet fields acquire very large VEVs. To determine them we consider
the $F$- and $D$-flatness conditions. Taking account of the  ${\cal Z}_2$ monodromy, $t_1\leftrightarrow t _2$  the $D$-flatness conditions are of the form given in Eq. (\ref{Dflat}) where there are three $U_{A}(1)$s with charges given in Eq.~(\ref{charges}). We wish to show that the D-flatness conditions are satisfied by the massless fields $\theta_{31},\;\theta_{34},\;\theta_{53}$ needed to give mass to exotics and, as discussed below, to generate  viable neutrino masses. Using the spectrum given in Table \ref{2} we compute $TrQ^{A}$ for the three $U_{A}(1)$s.  In a general basis, $Q = {\rm diag}[t_1, t_2, t_3, t_4, t_5]$, Eq. (\ref{Dflat}) can be written
\begin{equation}
(t_5 - t_3)|\theta_{53}|^2+(t_3 - t_4)|\theta_{34}|^2+(t_3 - t_1)|\theta_{31}|^2 = -X TrQ^A
\end{equation}
\noindent The trace is taken over all states, and is given by
\begin{equation}
Tr Q^A = 5 \sum n_{ij}(t_i + t_j)+10 \sum n_{k} t_k + \sum m_{ij}(t_i - t_j)
\end{equation}
\noindent For our model, this trace is computed to be
\begin{equation}
Tr Q^A = 61 t_1 -26 t_3 +14 t_4 +11 t_5
\end{equation}
\noindent Applying this to the three $U_{A}(1)$s using the generators given in Eq. (\ref{charges}) leads to

\bea
5|\theta_{53}|^{2}&=&5\,X\;\;(Q_{\chi})\nonumber\\
-|\theta_{53}|^{2}+4|\theta_{34}|^{2}&=&7\,X\;\;(Q_{\psi})\nn\\
2|\theta_{53}|^{2}-2|\theta_{34}|^{2}-3|\theta_{31}|^{2}&=&-113\,X\;\;(Q_{\perp})
\label{D1}
\eea
where $X=\frac{g_{s}^{2}M_{S}^{2}}{192\pi^{2}}$. These equations are solved by

\bea
|\theta_{53}|^{2}&=&X\nn\\
|\theta_{34}|^{2}&=&2\,X\nn\\
|\theta_{31}|^{2}&=&37\,X
\label{D2}
\eea

It remains to demonstrate $F$-flatness. The only allowed superpotential terms that can give a non-zero $F$-term involves the fields with VEVs plus at most a single additional light field. The only problematic terms have the form
$\lambda_{ij}\theta_{53}\theta^{i}_{31}\theta^{j}_{15}$ where $i=1,2,3,4$ and $j=1,2,3$. The F-terms of $\theta^{j}_{15}$ are potentially non-zero but minimisation of the singlet potential will make $\lambda_{i1}\langle\theta^{i}_{31}\rangle=0$ and $\lambda_{i2}\langle\theta^{i}_{31}\rangle=0$. This means three independent $\theta^{i}_{31}$ fields have zero VEVs but the fourth
one can have a VEV as it decouples from $\theta^{j}_{15}$. It is this combination that enters in Eqs. (\ref{D1}) and (\ref{D2}).

To complete the singlet discussion we note that most of the singlets acquire mass through the singlet VEVs. In particular the coupling $\theta_{14} \theta_{45} \theta_{53}\theta_{31}/M$
is allowed by the symmetries and generated a high scale mass, $\langle\theta_{53}\theta_{31}\rangle/M$ for the $\theta_{14}$ and $\theta_{45}$ fields. The reason these vector-like sets of singlet fields were included in the spectrum was to ensure the D-flat conditions could be satisfied - they play no role in the low energy phenomenology.
Since we have three $\theta_{14}$ fields and two $\theta_{45}$ fields in the massless spectrum below the string scale there will be one field $\theta_{14}$ left over.

{\subsection{Gauge coupling unification}\label{unification}

As we have seen, the result of flux splitting is to add SM non-singlet states in vector-like representations. Although these all acquire a large mass they still affect gauge coupling running. If they come in complete $SU(5)$ representations they do not affect the relative gauge coupling running at the one loop order. However this is not the case as may be seen from Table \ref{2} where there are incomplete multiplets. Therefore we would not
expect the gauge couplings to meet at some unification scale $M_{GUT}$.
On the other hand, as discussed in Appendix~\ref{B}, in F-theory
it has been observed~\cite{Blumenhagen:2008aw} that $U(1)_Y$ flux mechanism
splits the gauge couplings at the unification scale. Taking into account flux threshold effects,
at the GUT scale the gauge couplings $\alpha_i =g_i^2/4\pi$ are found to satisfy the relation,
\be
\frac{1}{\alpha_Y(M_{GUT})}=\frac 53 \,\frac{1}{\alpha_1(M_{GUT})}=\frac{1}{\alpha_2(M_{GUT})}+\frac 23 \frac{1}{\alpha_3(M_{GUT})}\label{SR},
\ee
Although this is not sufficient to yield a prediction for the low energy gauge couplings,
we find that the spectrum of exotics in the considered model tends to compensate
for the flux splitting at $M_{GUT}$,
so that the low energy gauge couplings can remain close to the same
values as predicted in conventional GUT models, independently of the exotic mass scale.
We shall also find that this exotic mass scale independence
remains true even when
two different exotic mass thresholds are taken into account.
The compensation is not exact however as noted in
\cite{Blumenhagen:2009yv} and in \cite{Dolan:2011iu} for the case of models with one or two $U(1)_{\perp}$s,
and we shall see that this also applies in the model with three $U(1)_{\perp}$s. Moreover,
the situation here is more involved because the exotic states are not expected to be degenerate,
so this requires a dedicated analysis for this particular model.

In our model we have the following vector pairs of exotics, which get large masses when $\theta_{31}$ gets a VEV according to Eq. (\ref{31}): ($Q+ \ov{Q}$), 2($u^c + \ov{u}^c$), ($d+ \ov{d}^c$), 2($L+ \ov{L}$), ($H_d + \ov{H}_d$).  Below some scale $M_X < M_{GUT}$ these exotics decouple.  We also have 3($D+ \ov{D}$), 2($H_u , H_d$) exotics which get masses when $\theta_{34}$ gets a VEV according to Eq. (\ref{34}).  Below a scale $M_{X \pr} < M_{X}$ these exotics decouple and only the MSSM spectrum remains massless for scales $\mu<M_{X \pr}$.
 As discussed above, the VEV for $\theta_{31}$ is expected to be much larger than that of $\theta_{34}$. Consequently the exotic states getting mass from different VEVs will have significantly different masses.
From Eqs. (\ref{31}) and (\ref{34}) we see that a vectorlike pair of $D,\;\bar{D}$ quarks in an incomplete $SU(5)$ multiplet are much lighter than the net $2L,\;2\bar{L}$ incomplete multiplets.
We find that this effect goes in the direction of cancelling the flux contribution. Since the effect depends sensitively on the scale at which gauge coupling running ceases, we now investigate the implications of these scales using
a one-loop renormalisation group analysis.

Assuming that extra matter decouples at mass scales $M_X$ and $M_{X \pr}$ ($M_Z < M_{X \pr} < M_X < M_{GUT}$) we can express the GUT scale as follows
(see Appendix~\ref{B} for details)
\begin{align}
M_{GUT} &=  e^{\frac{2\pi}{\beta {\cal A}}\rho}\, M_Z^{\rho} M_{X \pr}^{\gamma - \rho} M_X^{1-\gamma}
\label{MU} \\
\rho  &= \frac{\beta}{\beta_x} \notag \\
\gamma &= \frac{\beta_{x \pr}}{\beta_x} \notag
M_Z
\end{align}
where ${\cal A}$ is a function of the experimentally known low energy values of the
SM gauge coupling constants
\ba
\frac{1}{ {\cal A}} &=& \frac 53 \,\frac{1}{\alpha_1(M_Z)}-\frac{1}{\alpha_2(M_Z)}-\frac 23 \frac{1}{\alpha_3(M_Z)}
\ea
where $\beta,\beta_{x \pr} , \beta_{x}$ are the beta-function combinations in the regions $M_Z < \mu < M_{X \pr}$, $M_{X \pr} < \mu < M_{X}$ and $M_X < \mu < M_{GUT}$ respectively
\ba
\beta_x&=b_Y^x-b_2^x-\frac 23b_3^x \label{betax}\\
\beta_{x \pr}&=b_Y^{x \pr}-b_2^{x \pr}-\frac 23b_3^{x \pr} \label{betaxpr}\\
\beta&=b_Y-b_2-\frac 23b_3 \label{beta0}.
\ea
Imposing the well known condition $c_1({\cal L})^2=-2$ which eliminates the exotic states
$(3,2)_{-5/6}+(\bar 3,2)_{5/6}$
originating from the $SU(5)$-adjoint  decomposition in the bulk, we note that
all other types of these extra states descend  from  $\Sigma_{10},\Sigma_5$
matter curves. Let us denote with
\[n_Q, n_{u^c}, n_{d^c},  n_L, n_{e^c}, n_h\]
the multiplicities of  all types of possible extra states, in a self explanatory notation.
 We find
\[ \beta_{x \pr} - \beta =-2\, n_Q+n_{u^c}+n_{e^c}\]
and a similar equation for the exotics which are in the massless spectrum above the $M_X$ scale. We observe that the above difference depends on
 the number of additional quark doublets, u-type
right handed quarks and electrons  $n_Q, n_{u^c}, n_{e^c}$ and is independent
of the number of other types of additional states.

In the model under consideration, we analyse this information for the spectrum with two decoupling scales in  Appendix~\ref{B}, where we find the remarkable result that the GUT scale
becomes independent of the decoupling scales $M_X$, $M_{X'}$,
\be
M_{GUT} =  e^{\frac{2\pi}{\beta { \cal A}}}\,
M_Z=e^{\frac{\pi}{6 { \cal A}}}\,
M_Z\approx 2\times 10^{16}\,{\rm GeV}
\ee
that is, it is identical with the MSSM GUT scale.
Although one cannot predict the low energy QCD coupling constant in
F-theory, in Appendix~\ref{B} we obtain an approximate lower bound
$\alpha_3 \ge 0.11$, which is consistent with experiment.

\subsection{Baryon- and lepton-number violating terms}
As discussed above the R-parity violating superpotential couplings $u^cd^cd^c$, $Qd^cL$, $Le^cL$, $\kappa LH_u$ are not allowed because of the underlying $U(1)$ symmetries which play the role of R-parity.
Dimension 5 terms in the Lagrangian,
corresponding to the superpotential terms $QQQL$ and $u^{c}u^{c}d^{c}e^{c}$,
which would be allowed by usual R-parity, are forbidden by the $U(1)$ symmetries that originate in the underlying $E_{6}$.

Of course one must be careful that spontaneous symmetry breaking terms coming from SM singlet field VEVs do not allow these dangerous operators to appear.  Allowing for arbitrary singlet fields to acquire VEVs the dangerous the baryon- and lepton-number violating operators arise through the terms  $\theta_{15}LH_{u}$, $(\theta_{31}\theta_{45}+\theta_{41}\theta_{35})10_M\overline{5_3}^2$ and $\theta_{31}\theta_{41}10_M^3\overline{5_3}$. Thus, provided $\theta_{15}$, $\theta_{41}$ and $\theta_{45}$ do {\it not} acquire VEVs these dangerous terms will not arise.

However this is not sufficient to ensure the absence of baryon and lepton number violating terms because, even in the absence of these VEVs, tree level graphs can generate the dangerous operators at higher order in the singlet fields.
The dangerous graph is shown in Fig.~\ref{PD} and is driven by colour triplet exchange coming from the couplings
\ba
10_M\,10_M\,5_{H_u}&\ra & QQD_h +\ldots  \nn\\
 5_{H_u}\bar 5_{\bar H_u}&\ra& M_DD_h\bar D_h +\ldots  \nn
 \\
 \theta_{34}5_1\bar 5_2&\ra&  \langle\theta_{34}\rangle\,D_h'\bar D_h{'''} +\ldots   =
 \langle\theta_{34}\rangle\,D\bar D +\ldots        \nn .
\ea

As may be seen from Table~\ref{2} only the states $D'_{h}$ and $\bar{D}_{h}'''$ appear in the spectrum with mass generated by the singlet VEV $ \langle\theta_{34}\rangle$ which from Eq. (\ref{D2}) is
predicted to be somewhat below the GUT scale. Since the choice of fluxes in Table~\ref{2} eliminates
light colour triplet states $D_h$ in the low energy spectrum, arising from $5_{H_u}$, there is
no reason to expect any KK modes with the quantum numbers of $D_h$ below the string scale
since there is no ground state with the colour triplet quantum numbers of $D_h$ below the string scale.
Similarly the choice of fluxes in Table~\ref{2} eliminates
light colour triplet states $D_h{''}$ in the low energy spectrum, arising from $5_{4}$, so there is
no reason to expect any KK modes with the quantum numbers of $D_h{''}$ below the string scale.

\begin{figure}[!b]
\centering
\includegraphics[scale=.7,angle=0]{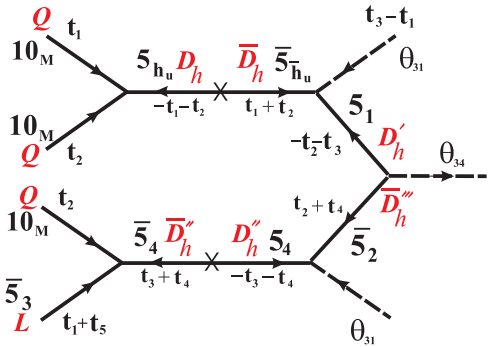}
\caption{\small{The proton decay diagram generating dim. 5 operator $QQQL$.}
} \label{PD}
\end{figure}

If string states with the quantum numbers of $D_{h},D_{h}''$ exist they are expected to have string scale masses, of $O(M_{S})$. In this case the diagram of Fig.~\ref{PD} gives the proton decay operator $QQQL$ with coefficient  $1/\Lambda_{eff}$ given by
\ba\frac{1}{\Lambda_{eff}}=\lambda^5\,\left(\frac{\langle\theta_{31}\rangle}{M_S}\right)^2
\frac{1}{\langle\theta_{34}\rangle}\label{lambda}
\ea
In (\ref{lambda}), $\lambda^5$ represents the the product of the five
Yukawa couplings in the relevant diagram and according to ref~\cite{Leontaris:2010zd}
it is expected to be
\[\lambda^5=\lambda_{10\cdot 10\cdot 5}\lambda_{10\cdot \bar 5\cdot\bar 5}\lambda_{5\cdot\bar 5\cdot 1}^3\approx 10^{-3}.\]

We can further determine the mass ratios taking into account the solution Eq. (\ref{D2}) to flatness conditions to estimate the effective scale
\[\Lambda_{eff} \approx 10^3\,\left(\frac{M_S}{\langle\theta_{31}\rangle}\right)^2
\frac{\langle\theta_{34}\rangle}{M_S}\, M_S\approx \frac{8\sqrt{6}\,\pi}{37g_s}\times 10^3 \,M_S\gtrsim 10^3\,M_S .\]

This, multiplied by the appropriate loop-factor due to higgsino/gaugino dressing
and other theoretical factors~\cite{Murayama:1994tc,Babu:1998wi,Goto:1998qg,Dermisek:2000hr,Murayama:2001ur},
 should be compared to experimental bounds on nucleon decay. This bound, relevant to the case that the operator $QQQL$ involves quarks from the two  lighter generations only,  requires $\Lambda_{eff}^{light}>(10^{8}-10^{9})M_{S}$.
 Given the large discrepancy between  $\Lambda_{eff}^{light}$ and $\Lambda_{eff}$ it is clearly important to determine whether, in the absence of flux, this light quark operator is generated by the diagram of Figure 1.

 Consider first trilinear couplings involving light fields only. They are given by an integral over the coordinates about the point of intersection, $z_{i}$, of the surface on which the matter curves reside. For the case there are $N$ multiple fields associated with a matter curve the orthogonal wave functions may be chosen proportional to powers of the coordinates, $(z_{i})^{j}$, $j=1,..,N$.  On integration only the coupling involving the fields with $j=0$ are non-zero, corresponding to a $U(1)_{i}$ invariance,  $z_{i}\rightarrow z_{i}e^{i\alpha_{i}}$. For the case the three families live on the same matter curve this means the mass matrices are rank 1 in the absence of flux. Switching on the flux gives a rank 3 mass matrix and generates the mixing between the generations.

 Now consider the case that there are vertices involving both light and heavy fields. In this case, because the heavy field wave function can involve powers of $\bar{z}_{i}$ \cite{Aparicio:2011jx}, there can be couplings involving light states with $j\ne 0$. However, as the
$U(1)_{i}$ invariance is intact, higher order operators with only external light fields are generated only if all the external fields have $j=0$.  In the absence of flux and assuming that the $\theta$ fields that acquire large vevs are also light fields this means that the operator generated by Fig 1 does not involve the light quarks. Thus its contribution to nucleon decay vanishes in the absence of flux and hence is significantly suppressed. To estimate this suppression we use the fact that the same flux effects generate the masses and mixings of the light quarks. Using these mixing angles we can convert the heavy quark operator to one involving light quarks. For the least suppressed case involving two down quarks and an up quark, this gives  $\Lambda_{eff}^{light}\approx \sqrt{\frac{m_{t}}{m_{u}}}\frac{m_{b}}{m_{d}}\Lambda_{eff}\approx 10^{9}M_{S}$, consistent with the experimental bound. A similar result applies to the operator involving right handed quarks.

\subsection{The $\mu$ term}
From Table \ref{2} it is clear that the $U(1)$ symmetries discussed above forbid a $\mu$ term. We expect these local symmetries to be anomalous and the associated gauge bosons to become massive due to the Stueckelberg mechanism, leaving three global $U(1)$ symmetries which act as selection rules in determining the allowed Yukawa couplings \cite{Grimm:2010ez}. However these global symmetries are only approximate and are spontaneously broken by the VEVs in Eq.(\ref{D2}). The $U(1)$ symmetries are also explicitly broken by
non-perturbative effects~\cite{Anderson:2009sw} with breaking characterised by the K\"ahler moduli, $\tau_{i}$, components of the  complex fields $T_{i}$, whose complex components provide the longitudinal components of the $U(1)$ gauge bosons. These non-perturbative effects will generate an explicit $\mu H_{u}H_{d}$ term with the $\mu = O(M_{s}e^{-t/M_{s}} )$ where $t$ is the VEV of the appropriate combination of $\tau_{i}$ moduli, and $M_s$ represents the string scale. Due to the exponential dependence on $t$ this term can be of the electroweak scale as required. Of course it is important that such breaking effects do not re-introduce nucleon decay terms at an unacceptable level. Provided the breaking of all the $U(1)$ symmetries are of the same order this will be the case because each of the dangerous operators will be suppressed by the factor $e^{-t/M}=O(\mu/M_{s})$. Thus the nucleon decay amplitude due to these terms will be suppressed by two powers of the string scale and be negligible.

 \subsection{Quark and charged lepton masses}
Up to SM singlets the surviving low energy spectrum is that of the MSSM given by:
\bea
&&\left[\overline{5}_3\rightarrow 3d^c+3L, \
10_M\rightarrow 3Q+3u^c+3e^c \right ], \nonumber \\
&&\left[\overline{5}_2\rightarrow H_d\right]_{10_{t_1}}, \nonumber \\
&&\left[5_{H_u}\rightarrow H_u\right]_{10_{t_3}}. \nonumber \\
\label{mssm}
\eea
The allowed low energy couplings in the superpotential originate from:
\bea
27_{t_1}27_{t_1}27_{t_3}&\rightarrow& 16_{t_1}16_{t_1}10_{t_3}
\nonumber \\
&\rightarrow& 10_M10_M5_{H_u}+ \overline{5}_3\theta_{15}5_{H_u}+
\overline{5}_310_2\overline{5}_2
\nonumber \\
&\rightarrow&
(3Q)(3u^c)H_u+(3L)(3\nu^c)H_u.
\label{yuk3}
\eea

A $3\times 3$ up-type and Dirac neutrino mass matrix is allowed at dimension three. In the absence of flux these matrices are rank one. However, as recently shown by Aparicio, Font, Ibanez and Marchesano~\cite{Aparicio:2011jx},  nonperturbative flux effects can generate an acceptable pattern for the light up quarks.

The down quark and charged leptons acquire mass through the non-renormalisable Yukawa couplings:
\bea
\theta_{31}27_{t_1}27_{t_1}27_{t_1}/M&\rightarrow& \theta_{31}16_{t_1}16_{t_1}10_{t_1}/M
\nonumber \\
&\rightarrow& \theta_{31}\overline{5}_310_M\overline{5}_2/M
\nonumber \\
&\rightarrow&
(\theta_{31}(3d^c)(3Q)+\theta_{31}(3L)(3e^c))H_d/M.
\label{dlyukawas}
\eea
Note that, from Table \ref{2},  the relevant graph \ref{botm} is generated by the exchange of a massive vectorlike pair  that is given a mass by $\langle\theta_{31}\rangle$.
\begin{figure}[!t]
\centering
\includegraphics[scale=.7,angle=0]{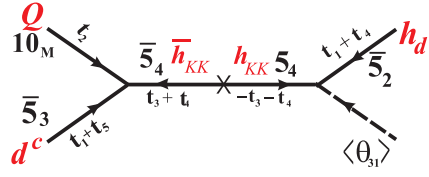}
\caption{\small{Tree-level diagram contributing to the bottom mass.}
} \label{botm}
\end{figure}
We already saw that $\theta_{31}$ must have a large VEV to give mass to exotics so this term can lead to down quark and charged lepton Yukawa couplings that are only mildly suppressed relative to the up quark couplings $\left(\langle\theta_{31}\rangle/M\ge m_{b}/m_{t}\right)$. This suppression provides an origin for the relative magnitude of the top quark to the bottom quark. Although the mass matrices for the down quarks and charged leptons coming from Eq. (\ref{dlyukawas}) are rank one, non-perturbative flux effects will generate the remaining terms and can lead to an acceptable mass structure~\cite{Aparicio:2011jx}.

\subsection{Neutrino masses}

As discussed in \cite{Bouchard:2009bu} models with a monodromy have states with Majorana mass. In this case, due to the monodromy, $\theta_{12}$ and $\theta_{21}$ are identified so, in the covering theory, the superpotential term $M_{M}\theta_{12}\theta_{21}$, that is allowed by all the symmetries, is a Majorana mass in the quotient theory. In what follows we shall use the notation $\Theta_{51}\equiv \theta_{53}\theta_{31}/M$. The RH neutrinos, $\theta^{i}_{15}$ couple to $\theta^{j}_{12}$ via $\lambda_{ij} \Theta_{51}\theta^{i}_{12}\theta^{j}_{15}$ where we have allowed for several states on the $\theta_{12}$ matter curve. In the absence of flux the mass mixing matrix $\lambda_{ij} \langle\Theta_{51}\rangle\theta^{i}_{12}\theta^{j}_{15}$ has rank one and we work in a basis in which only $\lambda_{11}$ is non-zero. With flux the remaining $\theta_{15}$ fields mix but this can be considerably suppressed. In the case of the up quark mass matrix such flux effects can explain the up quark mass hierarchy $m_{c}/m_{t}=10^{-2},\;m_{u}/m_{t}=10^{-4}$ so we may expect similar hierarchies in the mass mixing parameters $\lambda_{ij}$.

Since $\Theta_{51}$ has a VEV the fields $\theta^{j}_{15}$ acquire a Majorana mass, $M_{15}$, through the see-saw mechanism giving
\begin{equation}
M_{15}=\frac{\lambda^{2}\langle\Theta_{51}^{\dagger}\rangle^{2}M_{M}}{(|\lambda \langle\Theta_{51}\rangle|^{2}+|M_{M}|^{2})}.
\label{M15}
\end{equation}
where we have suppressed the family indices for clarity.  In the absence of flux only $\theta^{1}_{15}$ acquires a mass. With flux the remaining fields also get a Majorana mass but this can be significantly suppressed due to the expectation of smaller $\lambda$s. Since the light neutrinos couple to $\theta_{15}$ via the term $\lambda 'LH_{u}\theta_{15}$\footnote{$\lambda'$ is a matrix of couplings. In the absence of flux it is also rank 1.}  they, in turn, acquire Majorana masses given by
\be
M_{\nu}=\frac{\lambda '^{2}\langle H_{u}\rangle^{2}M_{15}}{(|\kappa\langle \Theta_{53}\rangle|^{2}+|M_{15}|^{2})}
\label{Mnu}
\ee
where the states $\theta_{15}$ also get mass from the coupling $\kappa\theta_{15}\theta_{31}\theta_{53}$.

What is the range of neutrino masses to be expected? From Eqs. (\ref{M15}) and (\ref{Mnu}) we have
\ba
M_{\nu}&<&\frac{\lambda'^{2}\langle H_{u}\rangle^{2}}{\lambda\langle\Theta_{51}\rangle}
\ea
For $\lambda,\;\lambda'=O(1)$, to get a mass of $O(10^{-1})eV$ requires $\langle\Theta_{51}\rangle\sim 10^{14}$ GeV which is too low given that ({\it c.f.} Eq. (\ref{D2})) $\langle\Theta_{51}\rangle/M\sim \theta_{31}/M\ge m_{b}/m_{t}$ and we expect $M\sim M_{S}$.

However it is clear the result is very sensitive to the couplings $\lambda$ and $\lambda'$ (and $\kappa$). Given the approximate rank 1 form of the matrix of couplings $\lambda$, from Eq. (\ref{M15}), two of the singlet states $\theta_{15}$ have suppressed mixing to $\theta_{12}$, which we characterise by $\tilde\lambda$, and consequently smaller Majorana mass. The resulting spectrum of doublet neutrino masses is one light one satisfying the bound of Eq. (\ref{Mnu}) due to the exchange of the heavy $\theta_{15}$ state plus two heavier states with mass satisfying the bound
\ba
M_{\nu}&<&\frac{\lambda'^{2}\langle H_{u}\rangle^{2}}{\tilde\lambda\langle\Theta_{51}\rangle}
\ea
due to the exchange of the lighter $\theta_{15}$ states. Note that in this equation we have kept the leading $\lambda'=O(1)$ coupling because we expect the light $\theta_{15}$ mass eigenstates states to contain a significant component of the state that has the leading $\lambda'$ coupling. As discussed above $\tilde\lambda$ can readily be of $O(10^{-2})$
or smaller. As a result, if this bound is saturated, two neutrino masses in the $10^{-1}eV$ range can readily be generated for $\langle\Theta_{51}\rangle/M_{S}\sim \theta_{31}/M\ge m_{b}/m_{t}$. However the saturation can be spoilt by the term involving $\kappa\langle\Theta_{53}\rangle$ in Eq. (\ref{Mnu}) so a determination of the precise result depends on the relative alignment of the leading contributions to $\lambda,\;\lambda'$ and $\kappa$. This in turn depends on the relative proximity of the relevant intersections of the matter curves involved in the three couplings.

\subsection{Relation to previous work}
In \cite{Ludeling:2011en} a general analysis was presented of the possible R-symmetries coming from the $U(1)_{\perp}$ factors in the local analysis of F-theory. Two possibilities were identified but it was shown that it was not possible to realise them in the semi-local picture. The model presented above corresponds to the Matter Parity Case 1 of \cite{Ludeling:2011en} and we have shown that it is consistent with the semi-local picture. The explanation of the apparent conflict is straightforward. In \cite{Ludeling:2011en}, seeking to generate viable fermion mass matrices without flux effects, the analysis considered only the case that the matter coming from the $10$ dimensional representation of $SU(5)$ should come from two matter curves, $10_{M}$ and $10_{t_{4}}$. As a result, in order to suppress the dimension 5 nucleon decay operators, a VEV for the field $\theta_{31}$ was forbidden and hence,  {\it c.f.} the discussion above, no down-type mass terms could be generated and the Matter Parity Case 1 was ruled out. However in the case of interest here all three generations are assigned to $10_{M}$. As a result  a VEV for $\theta_{31}$ is allowed without generating dimension 5 nucleon decay operators. Hence a down-type mass matrix proportional to $\langle \theta_{31}\rangle$
is possible and, allowing for flux effects, the resulting mass matrix can be of rank 3.

In \cite{Dolan:2011iu} a general discussion was presented of the difficulty in obtaining phenomenological viable F-theory models in the semi-local approach. The difficulty of reconciling the exotic spectrum necessitated by flux breaking with the $\mu$-term, the suppression of nucleon decay operators and gauge unification was emphasised and studied in detail for the case of models with one or two $U(1)_{\perp}$s. The model constructed here has three $U(1)_{\perp}$s and demonstrates that the problems can be ameliorated but not eliminated. In particular we have shown that the suppression of the dangerous nucleon decay operators is maintained while generating a $\mu$-term. However the constraints following from anomaly cancellation \cite{Dudas:2010zb,Marsano:2010sq} are still severe and lead to an extended exotic spectrum that affect gauge coupling running as discussed in Section \ref{unification}.

\section{Conclusions}

In the present work we  have  considered semi-local F-theory GUTs arising from a single $E_8$ point of local enhancement, involving simple GUT gauge groups based on $E_6$, $SO(10)$ and $SU(5)$
together with $SU(3)$, $SU(4)$ and $SU(5)$ spectral covers, respectively.
Assuming the minimal ${\cal Z}_2$ monodromy, we {determined} the homology classes of the spectrum for each case, and the implications for the resultant spectrum after flux breaking.  Our analysis includes singlets which have hitherto been ignored but which play a crucial role in phenomenology.

Using this, and aided by a dictionary relating the $E_{6}$, $SO(10)$, $SU(5)$ and singlet representations, we
constructed a model that leads to the MSSM at low energies. We showed that D-and F-flatness constraints require VEVs for singlet fields, which spontaneously break the global $U(1)$ symmetries,
and which generate large masses for all the non-MSSM exotic fields. In the absence of flux, the quark and charged lepton mass matrices are of rank one. When flux and instanton corrections are included, light quark and lepton masses and mixings are generated that can be consistent with their observed values. In the absence of flux, the additional $U(1)$ symmetries descending from $E_{8}$ ensure that dangerous baryon- and lepton-number violating terms are absent up to and including dimension 5, even taking into account the singlet VEVs which break the $U(1)$ symmetries. Including the flux effects, dimension 5 terms involving light quarks are generated but at an acceptable level. As a result the nucleon is stable within present limits without requiring super-Planckian messenger masses. The $\mu$ term in the theory is also forbidden by the $U(1)$ symmetries but can be generated at the SUSY breaking scale, again through non-perturbative effects which explicitly break the $U(1)$ symmetries.  Neutrino masses are generated via the see-saw mechanism, involving singlet neutrinos that acquire large Majorana masses allowed by the monodromy.

In conclusion, we have provided an example of a fully viable F-theory GUT, assuming flux breaking of all symmetries, satisfying the semi-local constraints, and employing only the additional (broken) $U(1)$ symmetries descending from the $E_{8}$ point of local enhancement.

\section*{Acknowledgements}
One of us (GGR) would like to thank Andre Lukas,  Luis Ibanez and Peter Nilles for useful discussions. JCC, SFK and GKL would like to acknowledge useful discussions with Q. Shafi at an early stage of this work. GKL would like to thank N.D. Vlachos for discussions. The research presented here was partially supported by the EU ITN grant UNILHC 237920 (Unification in the LHC era) and the ERC Advanced Grant BSMOXFORD 228169.

\newpage

\section*{Appendix}

\begin{appendix}

\section{\label{A} The homology classes of the Singlets}

  In order to determine the homology classes for the singlets of
 a particular  $SU(n)$ , we first  need to express the product of the
 the differences of the roots $t_i-t_j$ of the $n^{th}$ degree polynomial
$P_n=b_ks^{n-k}$ in terms of its coefficients $b_k$.

 Consider first the simplest case $b_0s^2+b_1s+b_2=0$.
 If $t_1,t_2$ are the roots, we know
 \[(t_1-t_2)(t_2-t_1)\propto -\Delta =-b_1^2+4b_0b_2\]
Note that  the same result is obtained from the determinant
\ba
\frac{1}{b_0}\left|
\begin{array}{ccc}
 b_0 & b_1 & b_2 \\
2 b_0 & 1 b_1 & 0 b_2\\
0 &2 b_0& 1 b_1
\end{array}
\right|= -b_1^2+4b_0b_2
\ea
We can easily repeat this procedure for the cubic roots.

 Consider now the generalization for the
$SU(4)$ case: According to known theorems
(see theorem 2.5 of ~\cite{Gelfand})
the required quantity is given by the
Sylvester formula
\ba
\left|
\begin{array}{ccccccc}
 b_0 & b_1 & b_2 & b_3 & b_4 & 0 & 0 \\
 0 & b_0 & b_1 & b_2 & b_3 & b_4 & 0 \\
 0 & 0 & b_0 & b_1 & b_2 & b_3 & b_4 \\
 4 b_0 & 3 b_1 & 2 b_2 & b_3 & 0 & 0 & 0 \\
 0 & 4 b_0 & 3 b_1 & 2 b_2 & b_3 & 0 & 0 \\
 0 & 0 & 4 b_0 & 3 b_1 & 2 b_2 & b_3 & 0 \\
 0 & 0 & 0 & 4 b_0 & 3 b_1 & 2 b_2 & b_3
\end{array}
\right|
\ea
If these are the roots of $SU(4)$, we have  $b_1=\sum_it_i=0$ and we get the result  (\ref{su4s}).

The extension to $SU(5)$ is straightforward. It can be computed form the determinant
\ba
\left|
\begin{array}{lllllllll}
 b_0 & b_1 & b_2 & b_3 & b_4 & b_5 & 0 & 0 & 0 \\
 0 & b_0 & b_1 & b_2 & b_3 & b_4 & b_5 & 0 & 0 \\
 0 & 0 & b_0 & b_1 & b_2 & b_3 & b_4 & b_5 & 0 \\
 0 & 0 & 0 & b_0 & b_1 & b_2 & b_3 & b_4 & b_5 \\
 5 b_0 & 4 b_1 & 3 b_2 & 2 b_3 & b_4 & 0 & 0 & 0 & 0 \\
 0 & 5 b_0 & 4 b_1 & 3 b_2 & 2 b_3 & b_4 & 0 & 0 & 0 \\
 0 & 0 & 5 b_0 & 4 b_1 & 3 b_2 & 2 b_3 & b_4 & 0 & 0 \\
 0 & 0 & 0 & 5 b_0 & 4 b_1 & 3 b_2 & 2 b_3 & b_4 & 0 \\
 0 & 0 & 0 & 0 & 5 b_0 & 4 b_1 & 3 b_2 & 2 b_3 & b_4
\end{array}
\right|
\ea
Setting $b_1=0$ we obtain the result quoted in the text.

\section{\label{B}RGEs and extra matter}

In this appendix we give a few details on the derivation of the GUT scale
and constraints on  other relevant quantities obtained from the renormalisation
group analysis.
It has been pointed out~\cite{Blumenhagen:2008aw}
 that the $U(1)_Y$ flux mechanism used to break the
$SU(5)$ gauge symmetry down to the Standard
Model one, splits the gauge couplings at the unification scale.
The splitting at $M_{GUT}$ is
\be\label{gcMU}
\begin{split}
\frac{1}{\alpha_3(M_G)}&=\frac{1}{\alpha_G}-y
\\
\frac{1}{\alpha_2(M_G)}&=\frac{1}{\alpha_G}-y+x
\\
\frac{1}{\alpha_1(M_G)}&=\frac{1}{\alpha_G}-y+\frac 35 x
\end{split}
\ee
 In the above we introduced the simplified notation $x=- \frac 12 {\rm Re} S\int c_1^2({\cal L}_Y)$ and  $y=\frac 12 {\rm Re} S\int c_1^2({\cal L}_a) $
associated with a non-trivial line bundle ${\cal L}_a$ and $S=e^{-\phi}+i\,C_0$ the
axion-dilaton field as discussed in~\cite{Blumenhagen:2008aw}.
Combining the above, the gauge couplings at $M_{GUT}$ are found to  satisfy  the  relation
\be
\frac{1}{\alpha_Y(M_{GUT})}=\frac 53 \,\frac{1}{\alpha_1(M_{GUT})}=\frac{1}{\alpha_2(M_{GUT})}+\frac 23 \frac{1}{\alpha_3(M_{GUT})}\label{SR2}
\ee
To obtain the low energy couplings, we use renormalisation group analysis at one-loop
level, taking into account threshold effects originating from possible existence
of exotic states appear in the spectrum. In general, different exotics decouple at
different scales.

In the considered model we have the following vector pairs of exotics, which get large masses when $\theta_{31}$ gets a VEV according to Eq. (\ref{31}): ($d+ \ov{d}^c$), ($Q+ \ov{Q}$), ($H_d + \ov{H}_d$), 2($L+ \ov{L}$), 2($u^c + \ov{u}^c$).  Below some scale $M_X < M_{GUT}$ these exotics decouple.  We also have 3($D+ \ov{D}$), 2($H_u , H_d$) exotics which get masses when $\theta_{34}$ gets a VEV according to to Eq. (\ref{34}).  Below a scale $M_{X \pr} < M_{X}$ these exotics decouple and only the MSSM spectrum remains massless for scales $\mu<M_{X \pr}$.
The low energy values of the  gauge couplings are then given by the evolution equations
\be\label{Brun}
\frac{1}{\alpha_a(M_Z)} = \frac{1}{\alpha_{a}(M_{GUT})}+\frac{b_a^x}{2\pi}\,\ln\frac{M_{GUT}}{M_X}+ \frac{b_a^{x\pr}}{2\pi}\,\ln\frac{M_{X}}{M_{X \pr}}+\frac{b_a}{2\pi}\,\ln\frac{M_{X \pr}}{M_Z}
\ee
where $b_a^x$ is the beta-function above the scale $M_X$,  $b_a^{x \pr}$ is the beta-function below $M_X$ and $b_a$ is the beta-function below $M_{X \pr}$.
Combining  the above equations, we find that the GUT scale is given by
\be
M_{GUT} =  e^{\frac{2\pi}{\beta {\cal A}}\rho}\, M_Z^{\rho} M_{X \pr}^{\gamma - \rho} M_X^{1-\gamma}
\label{M_U}
\ee
where ${\cal A}$ is a function of the experimentally known low energy values of the
SM gauge coupling constants
\ba
\frac{1}{\cal A} &=& \frac 53 \,\frac{1}{\alpha_1(M_Z)}-\frac{1}{\alpha_2(M_Z)}-\frac 23 \frac{1}{\alpha_3(M_Z)}
\nn\\
&=&\frac{\cos(2\theta_W)}{\alpha_{em}}-\frac 23 \frac{1}{\alpha_3(M_Z)}
\ea
where use has been made of the relations $\alpha_Y=\alpha_e/(1-\sin^2\theta_W)$ and $\alpha_2=\alpha_e/\sin^2\theta_W$.
We have also introduced  the ratios $\rho$ and $\gamma$
\be
\rho  = \frac{\beta}{\beta_x} \, \, \gamma = \frac{\beta_{x \pr}}{\beta_x}
\ee
where $\beta,\beta_{x \pr} , \beta_{x}$ are the beta-function combinations in the regions $M_Z < \mu < M_{X \pr}$, $M_{X \pr} < \mu < M_{X}$ and $M_X < \mu < M_{GUT}$ respectively
\begin{align}
\beta_x&=b_Y^x-b_2^x-\frac 23b_3^x \label{betax2}\\
\beta_{x \pr}&=b_Y^{x \pr}-b_2^{x \pr}-\frac 23b_3^{x \pr} \label{betaxpr2}\\
\beta&=b_Y-b_2-\frac 23b_3 \label{beta02}
\end{align}
Recall now the beta-function coefficients   ( $b_1=\frac 35\, b_Y$)
\ba
b_1&=&-0+2 n_G+\frac{3}{10}(n_h+n_L)+\frac{1}{5}n_{d^c}+\frac{1}{10}n_Q+\frac{4}{5}n_{u^c}
+\frac 35\,n_{e^c}
\\
b_2&=&-6+2n_G+\frac 12 (n_h+n_L)+0\,n_{d^c}+\frac 32\,n_Q+0\,n_{u^c}
\\
b_3&=&-9+2 n_G+0\,(n_h+n_L)+\frac 12\,n_{d^c}+n_Q+\frac 12\,n_{u^c}
\ea
with $n_G=3$ the number of families and $n_{h,L,...}$ counting  Higgses and extraneous matter.

Below $M_{X \pr}$ we have only the MSSM spectrum, thus $n_G=3,n_h=2$ and all
extra matter contributions are zero, $n_i=0$, thus
\ba
\{b_Y,b_2,b_3\}=\{11,1,-3\}&\ra&\beta =b_Y-b_2-\frac 23b_3=12\nn
\ea
In our model we have additional matter of 3($D+ \ov{D}$), 2($H_u , H_d$) above the scale $M_{X \pr}$.
Assuming $n_Q$, $n_{d^c}$   and $n_{u^c}$ extra $Q= (3,2)$, $d^c=(\bar 3,1)$ and $u^c=(\bar 3,1)$ and $n_h$ doublets
while writing $\beta_{x \pr}=\beta+\delta\beta_{x \pr}$  we have
\[ \delta\beta_{x \pr}=\beta^{x \pr}-\beta= -2\, n_Q+n_{u^c}\]
In our model
\[n_Q=0, n_{u^c}=0, n_{d^c}=6,  n_L+ n_h=4, n_{e^c}=0\]
thus
\[\delta\beta_{x \pr}=0\ra \beta_{x \pr}=\beta=12,\; \frac{\rho}{\gamma} =1\]

Above the scale $M_X$, we have additional matter ($d+ \ov{d}^c$), ($Q+ \ov{Q}$), ($H_d + \ov{H}_d$), 2($L+ \ov{L}$), 2($u^c + \ov{u}^c$).  As such
\[n_Q=2, n_{u^c}=4, n_{d^c}=2,  n_L+ n_h=6, n_{e^c}=0\]
and so
\
\[\delta\beta_{x}=\beta^{x}-\beta^{x \pr}=0\ra \beta_{x}=\beta=12,\; \rho = \gamma =1\]
From (\ref{M_U}) we see  that $M_{GUT}$ becomes independent of
the $M_X$ and $M_{X'}$ scales and in fact it is identified with the MSSM unification scale
\begin{equation}
M_U=M_{GUT}\equiv e^{\frac{2\pi}{\beta {\cal A}}}\,M_Z\approx 2\times 10^{16}\textrm{GeV}
\end{equation}

\subsection{A lower bound on the low energy QCD coupling constant}

Recall now that the parameter $x$ is given by  $x=-\frac{1}{2}\,{\rm Re} S\, \int c_1(L_Y)^2$,
with $S=e^{-\phi}+i\, C_0$ being the axion-dilaton.
Notice that elimination of unwanted exotics $(3,2)_{-5/6}+(\bar 3,2)_{5/6}$ (arising from
the  $SU(5)$ adjoint decomposition) impose the condition $\int c_1(L_Y)^2=-2$.
The low energy values of the  gauge couplings are then given by the evolution equations Eq. (\ref{Brun}). These imply
\begin{equation}
\left(\frac{1}{\alpha_2}-\frac{1}{\alpha_3}\right)_{M_Z}= x+ \frac{b_2^x-b_3^x}{2\pi}\log\left(\frac{M_G}{M_X}\right)
+ \frac{b_2^{x \pr}-b_3^{x \pr}}{2\pi}\log\left(\frac{M_X}{M_{X \pr}}\right) \frac{b_2-b_3}{2\pi}\log\left(\frac{M_{X \pr}}{M_Z}\right)
\end{equation}
\noindent Let us investigate the implications of the parameter $x$ which is found to
play the decisive role in the gauge coupling splitting. We note first
that a suitable twisting of ${\cal L}_a$ bundle by a trivial line bundle
${\cal L}_a\ra {\cal L}_a\times {\cal R}_a$ implies the following
change~\cite{Blumenhagen:2008aw}
\begin{equation}
\int c_1({\cal L}_Y)^2\;\ra\; \int c_1({\cal L}_Y)^2+2  c_1({\cal L}_Y)\cdot c_1({\cal L}_a)
\end{equation}
\noindent Since we have imposed the condition $\int c_1({\cal L}_Y)^2=-2$ and we assume
that the manifold is a del Pezzo surface $dP_n$, we conclude that
$\alpha_y=c_1({\cal L}_Y)$  is a root of the corresponding Lie Algebra. If now
the twisting is chosen so that $\alpha_a=c_1({\cal L}_a)$ is also a root of
the Lie Algebra, then $\alpha_Y\cdot \alpha_a=1$.  In this case $x=0$ and the
splitting effect vanishes since the remaining parameter $y$ induces only a
common shift to all gauge couplings, leading only to a redefinition
$\alpha_G^{-1}\ra \tilde \alpha_G^{-1}=\alpha_G^{-1}-y$. Thus, in this limiting case
we get the  standard gauge coupling unification scenario.

We proceed with the analysis for $x\ne 0$.
In our case the required beta functions combinations are
\ba
b_2 - b_3 &= 4 \\
b_2^{x \pr} - b_3^{x \pr} &= 4 \\
b_2^x - b_3^x &=3
\ea
Thus
\begin{equation}
x=\frac{1}{\alpha_2}-\frac{1}{\alpha_3}-\frac{2}{\pi}\ln\left(\frac{M_G}{M_Z}\right)-\frac{1}{2 \pi} \ln\left(\frac{M_{X \pr}}{M_{X}}\right)
\end{equation}

\noindent We have $\langle \theta_{31} \rangle = \sqrt{37 X}$ and
$\langle\theta_{34} \rangle = \sqrt{2 X}$ from Eq. (\ref{D2}), and so

\begin{equation}
\frac{M_{X \pr}}{M_X} = \sqrt{\frac{2}{37}}
\end{equation}

\noindent As such, we have

\begin{equation}
x=\frac 43\frac{1}{\alpha_2}-\frac 13\frac{1}{\alpha_Y}-\frac 79\frac{1}{\alpha_3}-\frac{1}{4 \pi} \ln\left({\frac{2}{37}}\right)
\end{equation}

Since $x$ is the dilaton field, $e^{-\phi}$ clearly we must have
 $x>0$ which gives a lower bound in $\alpha_3$
\begin{equation}
\alpha_3\ge \frac{7}{9}\frac{1}{\frac{5 \,{\sin^2\theta_W}-1}{3 \,\alpha_e}
-\frac{1}{4 \pi} \ln\left({\frac{2}{37}}\right)} \approx 0.1130
\end{equation}

\end{appendix}


\end{document}